\documentclass[prb,aps,nofootinbib, twocolumn, amssymb,superscriptaddress,longbibliography,10pt]{revtex4-2}
\usepackage{amsmath}
\usepackage{amssymb}
\usepackage{amsthm}
\usepackage{amsfonts}%
\usepackage{enumerate}
\usepackage{latexsym}
\usepackage{color}
\usepackage{setspace} 
\usepackage{blindtext}
\usepackage{dsfont}
\usepackage{mathrsfs}
\usepackage[normalem]{ulem}

\usepackage{tabularx}
\newcolumntype{Y}{>{\centering\arraybackslash}X}

\usepackage{stackengine}

\usepackage{bm}
\usepackage{graphicx}
\usepackage{subfigure}

\newcommand\xrowht[2][0]{\addstackgap[.5\dimexpr#2\relax]{\vphantom{#1}}}

\usepackage{hyperref}
\hypersetup{
pdfnewwindow=true, colorlinks=true,
linkcolor=blue, anchorcolor=blue,
citecolor=blue, filecolor=blue,
menucolor=blue, urlcolor=blue} 

\usepackage{pifont}

\newcommand{\vk}{\mathbf{k}}

\newcommand{\vq}{\mathbf{q}}

\begin{document}

\title{Kondo effect and its destruction in hetero-bilayer transition metal dichalcogenides}

\author{Fang Xie}
\thanks{\href{fx7@rice.edu}{fx7@rice.edu}}
\affiliation{Department of Physics \& Astronomy,  Rice Center for Quantum Materials, Rice University, Houston, Texas 77005, USA}
\affiliation{Rice Academy of Fellows, Rice University, Houston, Texas 77005, USA}
\author{Lei Chen}
\affiliation{Department of Physics \& Astronomy,  Rice Center for Quantum Materials, Rice University, Houston, Texas 77005, USA}
\author{Qimiao Si}
\affiliation{Department of Physics \& Astronomy,  Rice Center for Quantum Materials, Rice University, Houston, Texas 77005, USA}

\date{\today}

\begin{abstract}
    Moir\'e structures, along with line-graph-based $d$-electron systems, represent a setting to realize flat bands. 
    One form of the associated strong correlation physics is the Kondo effect. 
    Here, we address the recently observed Kondo-driven heavy fermion state and its destruction in AB-stacked hetero-bilayer transition metal dichalcogenides, which can be controlled by the gate voltages.
    By studying an effective interacting Hamiltonian using the slave spin approach, we obtained a phase diagram with the total filling factor and the displacement field strength as the tunable parameters. 
    In an extended range of the tunable displacement field, our numerical results show that the relative filling of the $d$ orbital, which is associated with the highest moir\'e band from the $\rm MoTe_2$ layer, is enforced to be $\nu_d \approx 1$ by the interaction. 
    This agrees with the experimental observation.
    We also argue that the observed high coherence temperature scale could be explained by the non-negligible bandwidth of the $d$ orbital.
    Our results set the stage to address the amplified quantum fluctuations that the Kondo effect may produce in these structures and new regimes that the systems open up for Kondo-destruction quantum criticality.
\end{abstract}

\maketitle

\section{Introduction}\label{sec:intro}

Moir\'e structures provide a setting to study strong correlation physics.
One of the first realized moir\'e structures is magic-angle twisted bilayer graphene, in which multiple strongly correlated phases have been observed in experiments \cite{cao_correlated_2018,cao_unconventional_2018}.
Multilayers of transition metal dichalcogenide (TMDC) represent another type of moir\'e structure to study correlation physics.
Pertinent to the TMDC moir\'e structure are a variety of strongly correlated phenomena, such as the correlated Chern insulator, the fractional Chern insulator, the Mott insulator, Wigner crystals, and the Kondo effect \cite{zhao2022gate,  zhao2023emergence,dalal_orbitally_2021,Kumar2022Gate,Guerci2023Chiral,Li2021Quantum,Li2021Continuous,ghiotto_quantum_2021,Zeng2023Integer,Park2023Observation, Cai2023Signatures,xu-observation-2023,wu_hubbard_2018,Zhang2021Spin,zang_2021_hartree,zang_dynamical_2022,Devakul2022Quantum,Dong2023Excitonic,morales-duran_nonlocal_2022,Pan2022Topological,Rademaker2022Spin,Yang2023Metal,Crepel2023Topological,Dong2023Composite,yu2023fractional}.

In these structures, the moir\'e energy bands near the Fermi level are typically narrow.
Therefore, the kinetic energy is reduced and the correlation effects are proportionally enhanced.
This type of behavior is also expected in other materials platforms, such as $d$-electron-based materials on kagome and other line-graph lattices with geometry-induced flat bands \cite{Ye2024, Huang2024, Ekahana2021.x}.
There has been a growing realization that these classes of systems can be described in terms of a Kondo lattice and the associated heavy fermion behavior, as arising in moir\'e structures based on TMDC \cite{zhao2022gate, zhao2023emergence,dalal_orbitally_2021,Kumar2022Gate,Guerci2023Chiral} and graphene \cite{Ram2021,Song2022Magic,zhou2023kondo,Hu2023Symmetric,Hu2023Kondo,dumitru2023twisted,huang_evolution_2023,Yu2023Magic,chou-kondo-2023,lau-topological-2023,li-topological-2023}, and in geometry-induced flat-band materials \cite{hu2023coupled, chen2022emergent, chen2023metallic}.
As such, these systems represent new platforms for emulating Kondo-driven correlation physics \cite{hewson1997kondo,Wirth2016,paschen_quantum_2021,Kirchner2020,Hu-qcm2022}.

Here, we show that the AB-stacked $\rm MoTe_2/WSe_2$ hetero-bilayer realizes a new correlation regime for both the Kondo effect and its destruction. 
To put our work into perspective, we now describe our specific motivations and provide a brief summary of the main points of our findings.

\begin{figure*}
    \centering
    \includegraphics[width=0.7\linewidth]{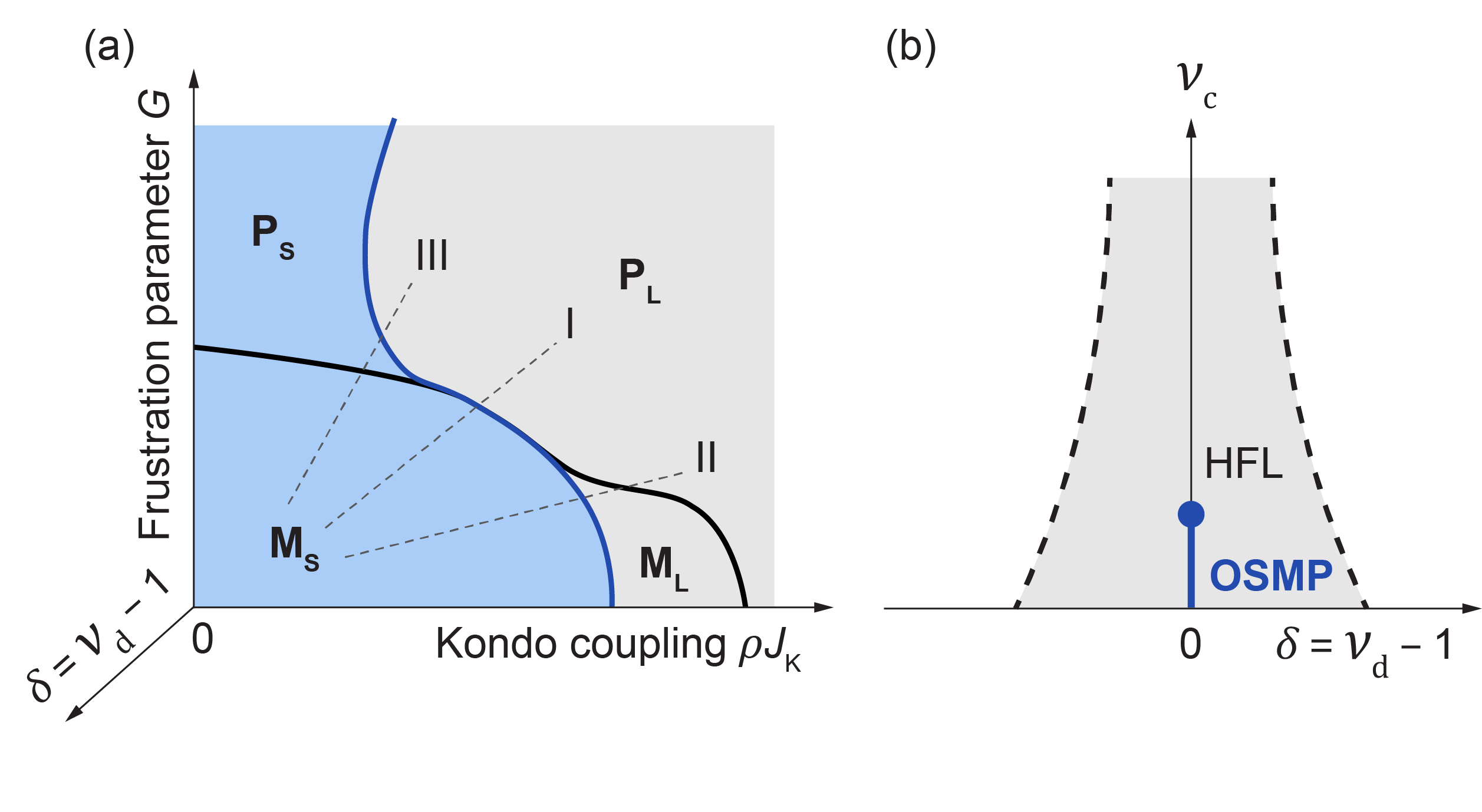}
    \caption{(a) The global schematic phase diagram of the Kondo lattice. Vertical axis $G$ stands for the frustration strength, and horizontal axis $\rho J_K$ describes the
    Kondo coupling. The three dashed lines represent possible quantum phase transition sequences between the Kondo-destroyed phases (light blue region: $M_S$, magnetic ordered small Fermi surface; or $P_S$, paramagnetic small Fermi surface) and the heavy Fermi liquid states (gray region: $P_L$, paramagnetic large Fermi surface; or $M_L$, magnetic ordered large Fermi surface). The third axis, which measures the deviation of the $d$ orbital filling from unity $\delta = \nu_d - 1$, is a new axis that becomes readily tunable in moir\'e heavy fermion systems. 
    (b) Schematic phase diagram of the two-orbital extended Hubbard model in the parameter space of 
    $d$ orbital filling factor $\delta = \nu_d - 1$ and $c$ orbital filling factor $\nu_c$. Here HFL stands for heavy Fermi liquid and OSMP (blue line) stands for orbital-selective Mott phase, in which the Kondo screening is destructed and local spin moments can form.}
    \label{fig:sketch}
\end{figure*}

\subsection{AB-stacked \texorpdfstring{$\rm MoTe_2/WSe_2$}{MoTe2/WSe2} hetero-bilayer}

In AB-stacked $\rm MoTe_2/WSe_2$ hetero-bilayer moir\'e structure, the two monolayers with lattice constants mismatch are stacked in opposite directions.
Moir\'e bands near the Fermi level from both layers form localized Wannier states sitting on hexagonal moir\'e superlattice sites \cite{Zhang2021Spin,Rademaker2022Spin} with different bandwidths, which allows a tight-binding description. 
More precisely, the moir\'e band which predominately comes from the $\rm MoTe_2$ layer has a smaller bandwidth than the other moir\'e band from the $\rm WSe_2$ layer. 
In the following, we will denote the Wannier orbital that predominately originates from the $\rm MoTe_2$ layer as $d$ orbital, and the orbital originating from the $\rm WSe_2$ layer as $c$ orbital.
Inter-orbital tunneling is strongly suppressed due to the spin-valley locking mechanism in TMDC materials \cite{Zhang2021Spin} and the opposite stacking directions. 
The inter-orbital tunneling is expected to be chiral due to the bilayer stacking \cite{Guerci2023Chiral}.
Because of the small bandwidths in moir\'e structures, the repulsive Coulomb interaction in the $c$ and $d$ orbitals is not negligible. Therefore, an extended two-orbital Hubbard model will be a reasonable minimum model. 

In transport experiments, the total electron filling factor and the displacement field perpendicular to the sample plane can be controlled directly by tuning the gate voltages. Since the $c$ and $d$ orbitals come from two different layers, the perpendicular displacement field induces a potential difference between these two orbitals, and thus, the electron fillings in the two individual orbitals can be controlled indirectly.

In this system, a heavy Fermi liquid state has been observed in an extended range of displacement field strengths \cite{zhao2022gate} (see Fig.~\ref{fig:exp-phase-diagram} in App.~\ref{appsec:exp}).
Quite noticeably, the coherence temperature controlling the crossover between the high-temperature incoherent scattering regime and the low-temperature Fermi liquid regime is very high: $T_{\rm coh}\approx 20\sim 40\,\rm K$ (called $T^*$ in Ref.~\cite{zhao2022gate}).
A sufficiently large magnetic field suppresses this heavy Fermi liquid state, as can be understood by the magnetic-field-induced suppression of the Kondo effect \cite{hewson1997kondo}.
Very recently, a magnetic ordered phase with destructed Kondo screening was also observed by reducing the filling of the $c$ orbital \cite{zhao2023emergence}.
These recent experimental discoveries indicate that the AB-stacked $\rm MoTe_2/WSe_2$ hetero-bilayer is a promising candidate for studying Kondo-driven correlation physics.

\subsection{Kondo effect}

The Kondo effect underlies the extremely strong correlations of heavy fermion metals.
Historically, heavy fermion metals have provided a canonical system for the understanding of Fermi liquid
phases with large renormalization factors \cite{hewson1997kondo}.

When the two bandwidths strongly differ, the narrower band can act as local moments 
while the wider band remains itinerant. This regime develops in frustrated $d$-electron systems, when the near-Fermi-energy flat bands and their topological nature are taken into account
\cite{hu2023coupled, chen2022emergent, chen2023metallic}, forming a new correlation regime in which the local Coulomb repulsion $U$ is in between the two bandwidths.

However, the bandwidths and the interaction of the AB-stacked $\rm MoTe_2/WSe_2$ hetero-bilayer
do not seem to fall in this regime. The ratio of the two bandwidths is only about $2$ and the interaction strength is not small in the $\rm WSe_2$ layer (see details below). 
One of the objectives of our work is to assess the correlation regime of the two-orbital Hubbard model. We anchor our consideration in terms of the coherent temperature, which has been experimentally measured, by exploring the implication of the aforementioned magnitudes of the bandwidths and interaction strength and the inter-layer hybridization strength. 
The observed coherent temperature is incompatible with any estimate in the local-moment-based
Kondo regime.
Using a saddle-point treatment of the microscopic Hamiltonian,  we argue that it instead implicates a more general correlation regime.

\subsection{Kondo destruction}

In the modern period of the heavy fermion field, heavy fermion metals have become active frontiers for the exploration of metallic quantum criticality \cite{Wirth2016,Kirchner2020,paschen_quantum_2021,Hu-qcm2022} (as well as strongly correlated metallic topology \cite{Lai2018,Dzsaber2017,Chen-Natphys22}).
A catalyst for extensive theoretical understanding and experimental studies has been the notion of Kondo destruction. 
The Kondo destruction theory was advanced in a study of the dynamical competition between the Kondo and Ruderman–Kittel–Kasuya–Yosida (RKKY) interactions in Kondo lattice models \cite{Si2001}, and was subsequently discussed \cite{Colemanetal} and studied \cite{senthil2004a} based on alternative approaches.
Experimental evidence has come from a variety of heavy fermion metals \cite{Schroder,paschen2004,shishido2005,park-nature06,Prochaska2020,Chen2023Shot}.
In the Kondo regime, the RKKY interaction favors a ground state with the quantum magnetism of the local moments. This process induces a dynamical competition against the formation of a Kondo singlet. The ensuing destruction of the Kondo effect leads to amplified quantum fluctuations. 
Because the nature of the ground state associated with the quantum magnetism itself can vary from genuine long-range magnetic order to being disordered (without symmetry breaking), the competition leads to a global phase diagram, as summarized in the main plane of Fig.\,\ref{fig:sketch}(a) \cite{{Si-physicab-06,Si_PSSB10,Coleman_Nev,Pixley-2014}}.
The Kondo lattice undergoes a transition from the heavy Fermi liquid state with a large Fermi surface ($P_L$ or $M_L$, labeled by gray), which incorporates the local moments into the Fermi volume, into a small Fermi surface state that does not count the local moments in the Fermi volume ($P_S$ or $M_S$, labeled by blue).
Most extensive studies have focused on the case when the ``$M$'' phase corresponds to an antiferromagnetic order, where the strength of frustration $G$ affects the degree of the order.
The case of ferromagnetic order has also been considered \cite{Yamamoto2010,Komijani2018}, though the role of frustration becomes uncertain.
The salient properties associated with the Kondo destruction QCP also includes a dynamical ``Planckian" ($\hbar \omega / k_{\rm B}T$) scaling, the associated linear-in-$T$ relaxation rate, and loss of the quasiparticles at the QCP \cite{Kirchner2020,paschen_quantum_2021,Hu-qcm2022,Si-JPSJ2014}.

For the AB-stacked $\rm MoTe_2/WSe_2$ hetero-bilayer, the extended correlation regime that has been implicated by considerations of the coherent temperature scale, as summarized in the previous subsection, also raises the question of the type of phase transition that can take place.
In this work, we address the issue within the saddle-point analysis in the non-magnetic sector.
Our analysis does find a Kondo destruction, starting from a regime of heavy fermion phase that is implicated by the magnitude of the observed coherence temperature.
Our results are qualitatively summarized in Fig.\,\ref{fig:sketch}(b). 
Our findings set the stage for further dynamical analyses of the model to explore the realization of other salient properties of the Kondo destruction. We discuss the issue further in Sec.\,\ref{sec:discussion}.

\subsection{A new axis in the phase diagram}

An important advantage of moir\'e structures is that a gate voltage readily tunes the electron density. 
The variation, however, is the total density $\nu$. We can express $\nu$ as the sum of the electron density for the $\rm MoTe_2$ layer, $\nu_d$, and that for the $\rm WSe_2$ layer, $\nu_c$. Thus, a new axis arises, namely the deviation of $\nu_d$ from half-filling. 
This is illustrated as a new axis of $\delta = \nu_d - 1$ in Fig.\,\ref{fig:sketch}(a) for the global phase diagram. More specifically, we illustrated the result of our saddle point calculations in the generalized phase diagram Fig.\,\ref{fig:sketch}(b), where the vertical axis represents a cut in the $\delta=0$ plane of the global phase diagram illustrated in Fig.\,\ref{fig:sketch}(a).
We stress two particular aspects of Fig.\,\ref{fig:sketch}(b). 
First, $\delta=0$ develops for a range of the total density $\nu$; this pinning of the $d$-electron density to half-filling results from strong correlations.
In this pinned range, the transition point between the heavy Fermi liquid and the orbital-selective Mott phase (OSMP) along the line captures an overall transition across the blue line in the global phase diagram of Fig.\,\ref{fig:sketch}(a). 
Second, the heavy Fermi liquid state can still exist if the $d$ orbital is slightly doped from unity (i.e., when $| \delta | \ll 1$). 
If the $d$ orbital filling factor $\nu_d$ is tuned significantly away from unity, one can anticipate a crossover out of the heavy Fermi liquid, resulting in a reduction in the effective mass of the $d$ orbital.

More specifically,
we use a saddle-point approach based on the U(1) slave spin representation \cite{Yu2012U1slave} to study the effective model for the TMDC hetero-bilayer with the total electron density $\nu = \nu_c + \nu_d$ and the displacement field potential $\varepsilon_D$ as tuning parameters. 
We demonstrate that the stripe-shaped heavy Fermi liquid region in this two-parameter space observed in experiment \cite{zhao2022gate, zhao2023emergence} can be well captured by our calculation.
The transition between the heavy Fermi liquid state and OSMP, which is associated with the reduction of the conduction electron density, is also observed in the numerical results shown in Fig.~\ref{fig:osmp}(d).

This article is organized as follows. In Sec.~\ref{sec:model}, we briefly introduce the tight-binding and interacting Hamiltonian and the definition of the tunable parameters that will be used later. Then we present the general and numerical results about the heavy Fermi liquid in Sec.~\ref{sec:hfl}, and about the orbital-selective Mott phase transition and Kondo destruction in Sec.~\ref{sec:osmp}. 
Finally we discuss results in Sec.~\ref{sec:discussion}, and we summarize our findings in Sec.~\ref{sec:summary}.

\section{Model}\label{sec:model}

In this section, we briefly introduce the model Hamiltonian for the AB-stacked bilayer $\rm MoTe_2/WSe_2$ system. We start from the effective tight-binding model on a hexagonal lattice, and then present an interacting Hamiltonian based on this lattice model. A detailed discussion about the Hamiltonian can be found in App.~\ref{appsec:model}.

\subsection{Tight-binding model}\label{sec:tightbinding}

\begin{figure*}[t]
    \centering
    \includegraphics[width=\linewidth]{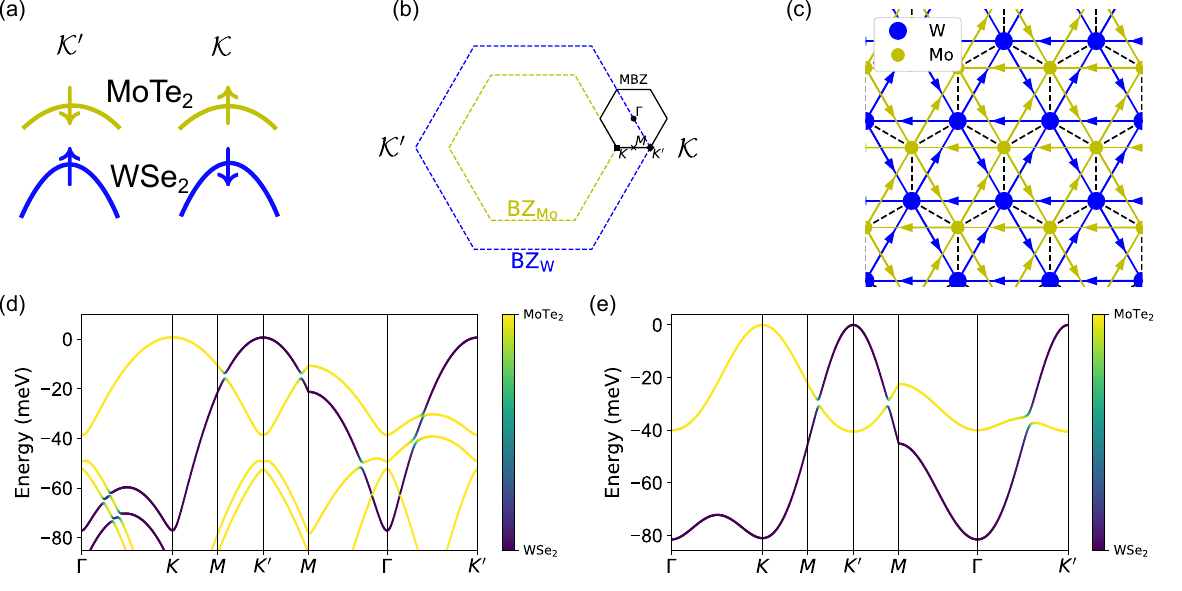}
    \caption{(a) The spin-valley locking of monolayer TMDC materials. Here we use $\mathcal{K}, \mathcal{K}'$ to represent the two valleys of the monolayer Brillouin zones. Since the two layers are stacked oppositely, the hole pockets in the same valley will have opposite spins. 
    (b) The relationship between the monolayer Brillouin zones (yellow and blue hexagons with dashed lines) and the moir\'e Brillouin zone (black hexagon). 
    (c) Effective tight-binding model of a single valley on a hexagonal lattice. Intra-layer hoppings in valley $\tau$ along the direction of the arrows have a phase factor $e^{\tau i \frac{2\pi}{3}}$ in the $\rm WSe_2$ layer, or $e^{-\tau i \frac{2\pi}{3}}$ in the $\rm MoTe_2$ layer. Inter-layer hoppings are shown by black dashed lines. Because of the spin-valley locking, these inter-layer hoppings flip the electron spin. 
    (d) The single valley band structure of the continuum model. The inter-layer hoppings open a small gap along $M$-$K'$-$M$ lines, and the highest moir\'e band above this hybridization gap carries non-zero Chern number $\mathcal{C} = \pm 1$.
    (e) The tight-binding model band structure with $t_c = 9~\rm meV$, $t_d=4.5~\rm meV$, and $t_{cd} = 1.5~\rm meV$. In this figure, the interlayer potential $\varepsilon_D$ is chosen such that the top band edges of the two bands are at the same energy. The tight-binding model is able to capture the key features of the highest moir\'e bands qualitatively.}
    \label{fig:non_interacting_model}
\end{figure*}

Monolayer TMDC materials in the 2H structural phase are known to have strong spin-orbital coupling, which leads to spin-valley locking \cite{Liu2013Three,Kormanyos2015kp}. As depicted in Fig.~\ref{fig:non_interacting_model}(a), the hole pocket near valley $\mathcal{K}$ exclusively has a spin-$\uparrow$ band, whereas the pocket near $\mathcal{K}'$ only contains a spin-$\downarrow$ band.

In hetero-bilayer TMDC systems, the difference in unit cell size results in a lattice mismatch that can cause the formation of a moir\'e superlattice at zero-angle twisting.
As a result, the size of the moir\'e Brillouin zone is determined by the difference between the Brillouin zones of the two monolayers, as illustrated in Fig.~\ref{fig:non_interacting_model}(b). 
In valley $\mathcal{K}$, the top band edge of the $\rm MoTe_2/WSe_2$ band will be centered around the $K/K'$ point in the moir\'e Brillouin zone, respectively.
Since the two layers are AB-stacked, the hole bands from the two layers in the same valley will carry opposite spin quantum numbers. As a consequence, the inter-layer hoppings only consist of spin-flipping terms, which are anticipated to be weak.

Similar to the Bistritzer-MacDonald model \cite{Bistritzer2011Moire} for twisted bilayer graphene, a single valley continuum model can also be derived for the AB-stacked TMDC \cite{Zhang2021Spin,Rademaker2022Spin}. 
The band structure of the continuum model can be found in Fig.~\ref{fig:non_interacting_model}(d).
Since the hole pockets in the two types of monolayers have different effective mass, the highest moir\'e bands will also exhibit different bandwidths: the moir\'e band from $\rm MoTe_2$ has a bandwidth $D_{\rm Mo}\sim 40~\rm meV$, while the other layer has $D_{\rm W}\sim 80~\rm meV$. 

Although the continuum model is more accurate, it is challenging to study compared to a simplified tight-binding model.
By analyzing the symmetry of the continuum model \cite{Bradlyn2017Topological}, a tight-binding model, which is able to capture the major low energy features of this system, can be constructed from the continuum model \cite{Zhang2021Spin,Devakul2022Quantum,Rademaker2022Spin,Dong2023Excitonic}.
The charge centers of the orbitals from each layer can form triangular \emph{moir\'e superlattices}, and they are located at two distinct hexagonal sites belonging to two different Wyckoff positions, which are represented by yellow and blue markers in Fig.~\ref{fig:non_interacting_model}(c). 
Therefore, the tight-binding model can be written on the hexagonal lattice, with two orbitals per unit cell. 
We use $c^\dagger_{i\tau}/d^\dagger_{i\tau}$ to represent the fermionic creation operator of the orbital from the $\rm WSe_2/MoTe_2$ layer and valley $\tau$ at superlattice site $i$. 
Here $\tau = \pm$ represent the single-layer valley indices, which correspond to $\mathcal{K}$ and $\mathcal{K}'$, respectively.
One can write down the following tight-binding Hamiltonian with four bands \cite{Devakul2022Quantum,Dong2023Excitonic}:
{\small
\begin{align}\label{eq:h0}
    H_0 &= \sum_{\langle\langle ij\rangle\rangle,\tau}t_{c}e^{\tau i\phi^c_{ij}}c^\dagger_{i\tau}c_{j\tau} + \sum_{\langle\langle ij\rangle\rangle, \tau}t_de^{\tau i\phi^d_{ij}}d^\dagger_{i\tau}d_{j\tau}\nonumber\\
    & + \sum_{\langle ij\rangle,\tau} \left(t_{cd}c^\dagger_{i\tau}d_{j\tau} + {\rm h.c.}\right) + \frac{\varepsilon_D}{2}\sum_{i\tau}\left(c^\dagger_{i\tau}c_{i\tau} - d^\dagger_{i\tau}d_{i\tau}\right)\,.
\end{align}}
Here $\langle \langle ij\rangle\rangle$ denotes the next-nearest neighbor sites, which are always intra-layer terms, while $\langle ij \rangle$ represents the nearest neighbor sites, exclusively connecting sites from different layers.
The phase factors associated with the intra-layer hoppings are chosen to be $\phi^c_{ij} = 2\pi/3$ and $\phi^d_{ij} = -2\pi/3$ along the arrows in Fig.~\ref{fig:non_interacting_model}(c). 
These hopping phases will result in the top band edges of the $\rm MoTe_2/WSe_2$ bands being located at the $K/K'$ points in valley $\tau = +$.
As shown by the dashed lines in Fig.~\ref{fig:non_interacting_model}(c), the inter-layer hoppings are along the nearest neighbor directions instead of on-site. This leads to a chiral signature with the form of $k_x \pm i k_y$ near $K$ and $K'$ in the hybridization function, which is different from the momentum-independent hybridization in standard Anderson models, as has been studied in Ref.~\cite{Guerci2023Chiral}. $\varepsilon_D$ represents the potential difference between the two layers induced by the displacement field. 

By fitting the tight-binding model with the dispersion of the continuum model, these parameters can be set to the following values:
\begin{align}
    t_c & \approx 9~{\rm meV}\,,\nonumber\\
    t_d & \approx 4.5~{\rm meV}\,,\nonumber\\
    t_{cd} & \approx 1.5~\rm{meV}\,.
\end{align}
The dispersion of the tight-binding model is also shown in Fig.~\ref{fig:non_interacting_model}(e). Qualitative features of the highest moir\'e bands in the continuum model can be well-captured by this simple tight-binding model. Therefore, we will use these tight-binding parameters throughout this manuscript unless otherwise stated.

\subsection{Interacting Hamiltonian}

Kondo-driven physics is usually modeled by the Anderson lattice Hamiltonian.
However, it is crucial to assess whether the AB-stacked TMDC hetero-bilayer can be well described by the standard Anderson lattice Hamiltonian. 
Indeed, a few factors beyond the Anderson lattice model need to be considered.
The bandwidth ratio between the two bands from the $\rm MoTe_2$ and $\rm WSe_2$ layers is not very small: $D_{\rm Mo}/D_{\rm W} \sim 0.5$. Both $D_{\rm W}$ and $D_{\rm Mo}$ might be comparable to the strength of the on-site interaction in the hetero-bilayer TMDC moir\'e material. 
Hence, (i) the dispersion of the $d$ bands is not negligible, and (ii) the interaction in the conduction band has to be considered.
We also note that (iii) the non-local interaction terms might not be negligible when compared with the on-site interaction \cite{wu_hubbard_2018, morales-duran_nonlocal_2022}. 
As indicated by previous studies, these non-local interaction terms are able to affect the quasiparticle weight $Z$ \cite{Hassan2010slave}, and contribute to the stabilization of inter-valley excitonic order, leading to a quantum anomalous Hall effect in this system \cite{Dong2023Excitonic,Xie2022Topological}.
Therefore, we choose the multi-orbital extended Hubbard Hamiltonian as our minimum model, which is more general than the standard Anderson lattice Hamiltonian, though in the latter case the effect of a correlated conduction band has been discussed \cite{Schork1997Periodic}.
The interacting Hamiltonian can be written the following form:
\begin{align}\label{eq:Hint}
    H_I =& \frac{U}{2}\sum_{i,\alpha=c,d}\left(n_{i\alpha+} + n_{i\alpha-} - 1\right)^2 \nonumber\\
    & + V\sum_{\langle ij \rangle, \tau\tau'}\left(n_{ic\tau}-\frac12\right)\left(n_{jd\tau'} - \frac12\right)\nonumber\\
    & +V'\sum_{\langle\langle ij\rangle\rangle,\alpha\tau\tau'} \left(n_{i\alpha\tau} - \frac12\right)\left(n_{j\alpha\tau'} - \frac12\right)\,,
\end{align}
in which $n_{i\alpha \tau} = \alpha^\dagger_{i\tau}\alpha_{i\tau}\,,~\alpha=c{\rm ~or~}d$ is the electron density operator. Here we use $U$ to represent the on-site interaction, $V$ to represent the inter-orbital [nearest-neighbor (NN)] interaction, and $V'$ to represent the next-nearest-neighbor (NNN) interaction. The interaction strength in the lattice model can be estimated from the Wannier functions of the orbitals, and the on-site interaction $U$ is indeed comparable to the bandwidth of the wide band $D_{\rm W}$.

\section{Heavy Fermi liquid}\label{sec:hfl}

\begin{figure*}[t]
    \centering
    \includegraphics[width=\linewidth]{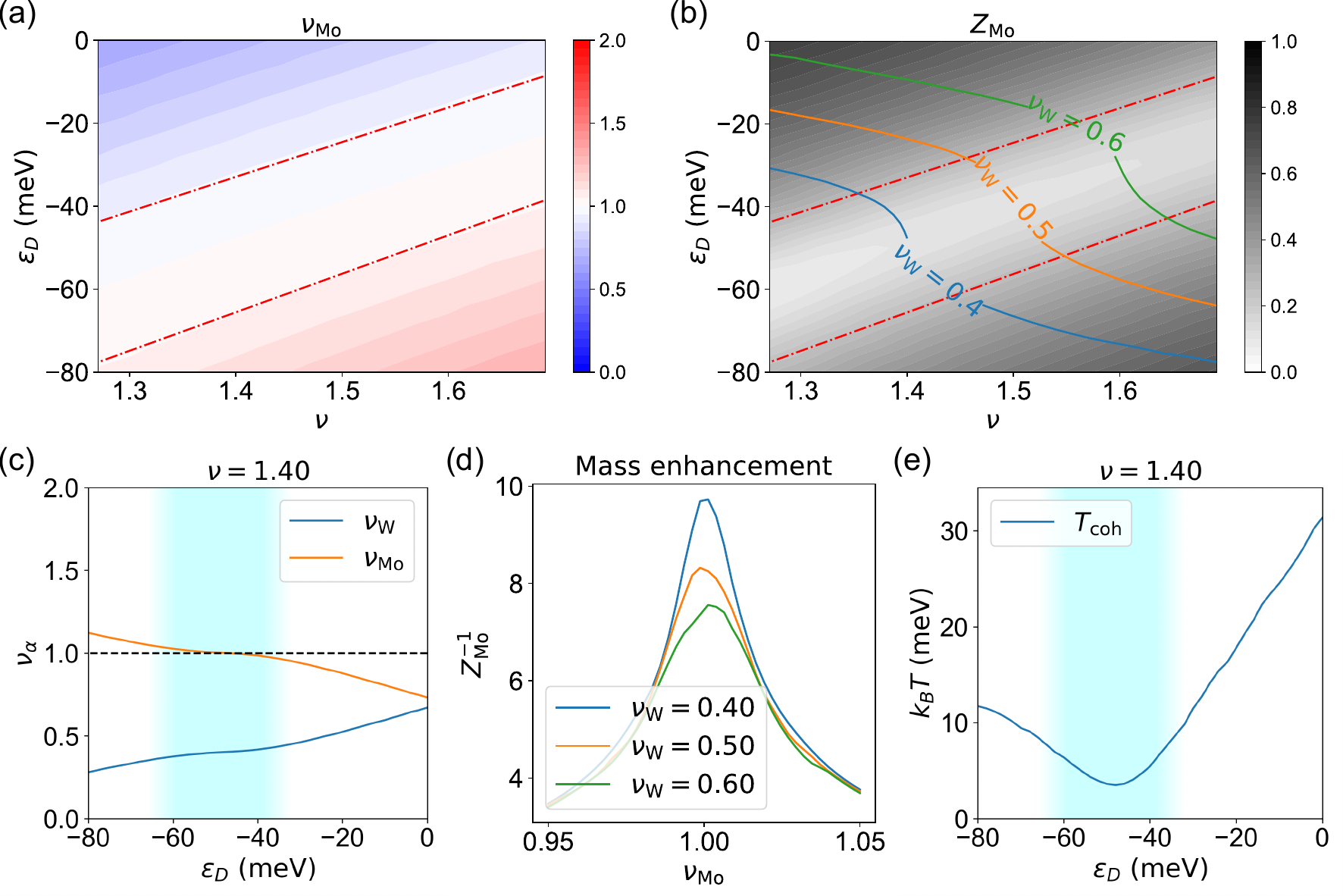}
    \caption{(a) The orbital resolved filling factor of the heavy $(\rm Mo)$ orbital $\nu_{\rm Mo}$ in the parameter space $(\nu, \varepsilon_D)$. In the region labeled by the red dashed lines, the $d$ orbital is almost at half filling $\nu_{\rm Mo} \approx 1$. (b) The quasiparticle weight of the $d$ orbital $Z_{\rm Mo}$ in the parameter space $(\nu, \varepsilon_D)$. The conduction-band filling factors $\nu_{\rm W}$ are constants along the three solid lines. (c) The filling factors of the two orbitals at a fixed total filling $\nu = 1.4$ with varying displacement field. The heavy Fermi liquid region is highlighted with light blue. (d) The mass enhancement (inverse of the quasiparticle weight $Z^{-1}$) of the $d$ orbital as a function of $d$ orbital filling, with multiple fixed conduction orbital fillings. (e) The coherence temperature scale evaluated with fixed total filling factor $\nu = 1.4$. We choose the strength of on-site interaction as $U = 70~\rm meV$, NN interaction $V = U/2$, and NNN interaction $V' = U/4$.}
    \label{fig:phase_diagrams}
\end{figure*}

\begin{figure}[t]
    \centering
    \includegraphics[width=\linewidth]{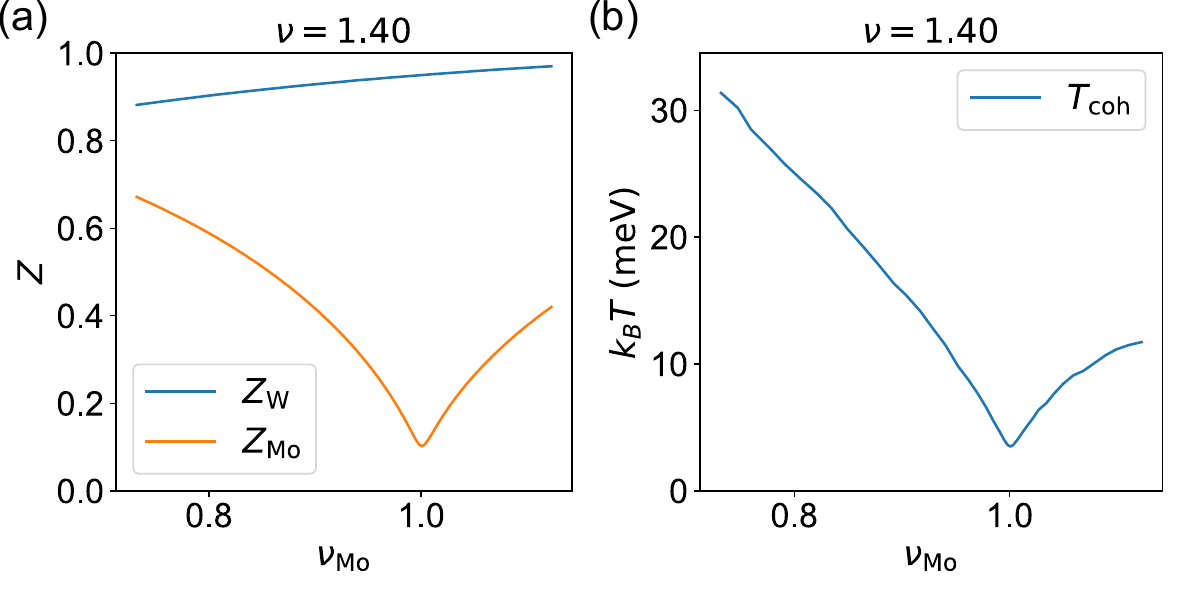}
    \caption{(a) The quasiparticle weights of both orbitals as functions of the heavy fermion filling factor $\nu_{\rm Mo}$. (b) The coherence temperature scale as a function of the heavy fermion filling factor. The total filling factor in both figures is fixed to be $\nu = 1.4$. The on-site interaction is chosen to be $U = 75\,\rm meV$ and $V = U/2$, $V' = U/4$.}
    \label{fig:nd-phase-diagram}
\end{figure}

Based on experimental observations, the heavy Fermi liquid behavior is predominantly observed within a stripe-like region in the $(\varepsilon_D, \nu)$ phase diagram \cite{zhao2022gate} when the total hole filling factor is in the interval $1 < \nu < 2$. 
Hence, we first choose the strength of on-site interaction as $U = 70\,\rm meV$, NN interaction as $V = U/2$, and NNN interaction as $V' = U/4$. 
We also consider different values of hole total filling factor $1.3 \lesssim \nu \lesssim 1.7$, and the displacement field in the range of $-80\,{\rm meV} \leq \varepsilon_D \leq 0\rm\,meV$. 
We proceed to solve the saddle-point equations in the U(1) slave spin approach, the details
of which are explained in App.~\ref{appsec:method}. 

\subsection{Development of heavy Fermi liquid}

The orbital resolved filling factors can be obtained from the U(1) slave spin approach, which are shown in Fig.~\ref{fig:phase_diagrams}(a). 
When the filling factor $\nu$ remains constant, there exists a range of displacement field strengths where the $\rm Mo$ layer orbitals are approximately pinned at unity filling ($\nu_{\rm Mo} \sim 1$). 
Remarkably, the width of this interval does not vary significantly with the total filling factor, resulting in a ``stripe'' region highlighted by the red dashed lines in Fig.~\ref{fig:phase_diagrams}(a), which resembles the experimental observation [see Fig.~\ref{fig:exp-phase-diagram} in App.~\ref{appsec:exp} for reference]. 
To demonstrate this clearly, we also provide the orbital resolved filling factors as functions of the displacement field potential $\varepsilon_D$ with fixed total filling factor $\nu = 1.4$ in Fig.~\ref{fig:phase_diagrams}(c). It is evident that the $d$ orbital ($\rm Mo$) filling exhibits a distinct ``plateau'' pattern when $\varepsilon_D$ is varied, corresponding to the stripe region in Fig.~\ref{fig:phase_diagrams}(a).
We also observed that the width of the stripe region in the phase diagram is around $\Delta \varepsilon_D \lesssim 30\,\rm meV$, which is clearly smaller than the on-site Hubbard interaction $U = 70\,\rm meV$. This could be attributed to the large bandwidth of the $d$ orbital.

Similar to the determination of orbital resolved filling factors, the evaluation of orbital resolved quasiparticle weight for the $d$ orbital $\rm Mo$ is also accessible through the slave spin
approach, the results of which are shown in Fig.~\ref{fig:phase_diagrams}(b).
This quasiparticle weight $Z_{\rm Mo}$ drops significantly to $Z_{\rm Mo} \lesssim 0.2$ within the heavy Fermi liquid region. 
To demonstrate how the $d$ orbitals are screened by the conduction electrons, we plot the mass enhancement of the heavy band, which is defined as the inverse of the quasiparticle weight, with different conduction electron densities in Fig.~\ref{fig:phase_diagrams}(d).
The effective mass always approaches its maximum value at $\nu_{\rm Mo} = 1$, and becomes smaller if the $d$ orbital is doped away from half-filling.
We also notice that the maximum value of the effective mass decreases with an increasing conduction electron density, because of the stronger screening. This picture is captured by the schematic phase diagram in Fig.~\ref{fig:sketch}(b).

The large mass enhancement for the $d$ orbital could potentially account for the experimentally observed narrow bandwidth of the $\rm Mo$ bands \cite{zhao2022gate}, whose corresponding effective mass $m^*_{\rm Mo} \approx 5\sim 10m_e$ is an order of magnitude larger than the value $m_{\rm Mo} \approx 0.65 m_e$ predicted by the non-interacting continuum model \cite{Zhang2021Spin,Rademaker2022Spin}. 
In contrast, the quasiparticle weight of the conduction electron [see Fig.~\ref{fig:add-Tcoh}(a) in App.~\ref{appsec:Tcoh}] remains relatively high with $Z \gtrsim 0.8$ across the entire parameter space examined. Consequently, the effective mass $m^*_{\rm W} \approx 0.5 m_e$ is similar to the value $m_{\rm W} \approx 0.35m_e$ predicted in the non-interacting theory.

\subsection{Coherence temperature scale}

The coherence temperature $T_{\rm coh}$, which characterizes the onset of Kondo screening as temperature is lowered, can be estimated from the effective bandwidth of the $d$ orbital in the zero temperature slave fermion Hamiltonian \cite{nozieres_comments_1998,burdin_coherence_2000,kourris_kondo_2023}. 
In practice, this ``bandwidth'' can also be estimated from the inverse of the local density of states of the slave fermion, which will be explained in detail in App.~\ref{appsec:Tcoh}.

Normally, one would expect that the Kondo screening temperature scale is on the order of $\frac{1}{2}D_{\rm W} \exp({-\frac{D_{\rm W}}{J_K}})$ \cite{hewson1997kondo}.
The Kondo coupling can be estimated by $J_K \sim t_{cd}^2/\Delta$ in which $t_{cd} \approx 1.5\,\rm meV$ is the hybridization, and $\Delta$ is the activation gap. Hence, the Kondo coupling strength is expected to be $J_K \lesssim 5\, \rm{meV}$, resulting in a very low screening temperature scale of $\lesssim 0.1\, \rm{K}$.
Alternatively, in Kondo lattice systems, the coherence temperature of a metallic Kondo screened phase is controlled by $T_{\rm coh} \sim r^2/D_{\rm W} \sim 0.05 \rm \, K$ \cite{burdin_coherence_2000}, in which the renormalized hybridization is $r \sim \sqrt{Z_{\rm Mo}} t_{cd} \sim 5 \rm\,K$.
Both estimations lead to notably low temperature scales.

However, the coherence temperature has been extracted from resistivity measurements \cite{zhao2022gate} to be much higher, $\approx 20 \sim 40\rm\,K$.
Indeed, the bandwidth of the $d$ orbital in hetero-bilayer TMDC is far from negligible, and thus the coherence temperature will depend on the quasiparticle weight $Z_{\rm Mo}$ in a different manner. 
As discussed in App.~\ref{appsec:Tcoh}, the coherence temperature can be estimated via the following expression when $D_{\rm Mo}$ is considered:
\begin{equation}
    T_{\rm coh} \sim Z_{\rm Mo}D_{\rm Mo} + \frac{Z_{\rm Mo}Z_{\rm W} t^2_{cd}}{D_{\rm W}}\,.
\end{equation}
Thus, much higher coherence temperatures become possible due to the dispersion of the heavy band. 
For fixed total filling factor $\nu = 1.4$, the value of $T_{\rm coh}$ as a function of displacement field can be found in Fig.~\ref{fig:phase_diagrams}(e).
In the heavy Fermi liquid region, the coherence temperature is estimated to be at the order of magnitude of $T_{\rm coh}\sim 50\, \rm K$. Notably, it aligns with the order of magnitude of $\sim 40\rm\,K$ measured in the experiment.
We also show that the quasiparticle weight of the $d$ orbital and the coherence temperature (Fig.~\ref{fig:nd-phase-diagram}) approach their minimum when the $d$ orbital is at half-filling.

\section{Orbital-selective Mott phase}\label{sec:osmp}

\begin{figure*}[!t]
    \includegraphics[width=\linewidth]{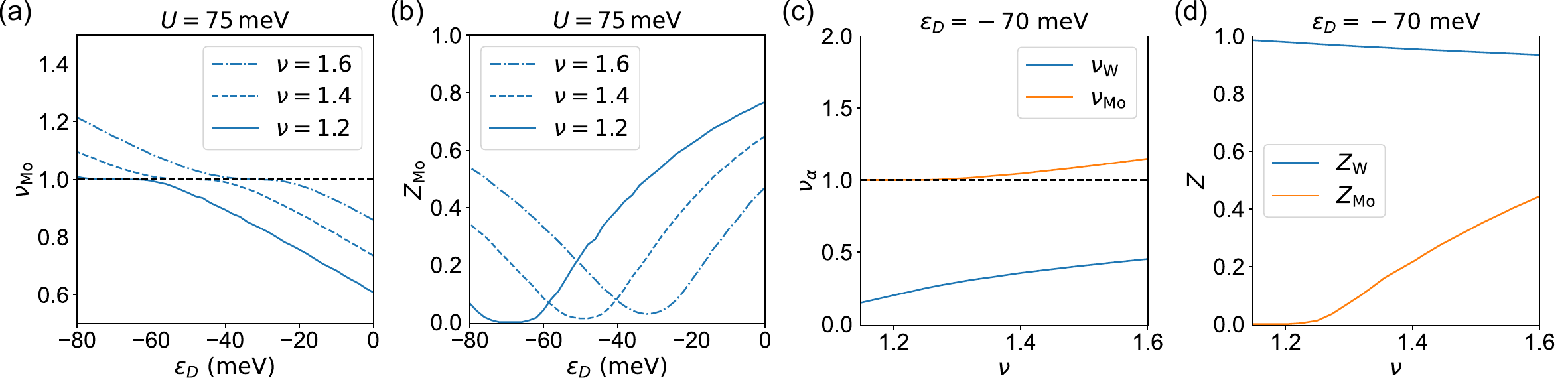}
    \caption{(a) The orbital resolved filling factors of the $\rm Mo$ orbital as a function of displacement field with different total filling factors. Data shown by dash-dotted line, dashed line and solid lines are obtained with $\nu = 1.6$, $\nu = 1.4$ and $\nu = 1.2$, respectively. 
    (b) The quasiparticle weight of the $\rm Mo$ orbital as a function of displacement field. The total filling factors are the same as in (a). 
    (c) The orbital resolved filling factors as functions of the total filling factor $\nu$, with fixed displacement field potential $\varepsilon_D = -70\,\rm meV$. It can be seen that the $d$ orbital $\rm Mo$ filling is almost pinned at $\nu_{\rm Mo} = 1$ when $\nu \lesssim 1.3$.
    (d) The quasiparticle weights as functions of the total filling factor $\nu$ with fixed displacement field. Orbital-selective Mott phase can be attained when $\nu \lesssim 1.2$.
    The on-site, NN, and NNN interactions are chosen to be $U = 75\rm\,meV$, $V = U/2$, and $V' = U/4$.}
    \label{fig:osmp}
\end{figure*}

An orbital-selective Mott phase (OSMP) develops when the $d$ electrons are localized while the $c$ electrons are itinerant.
The transition from the heavy Fermi liquid into the OSMP can be achieved by tuning the total filling factor. In Figs.~\ref{fig:osmp}(a) and \ref{fig:osmp}(b), we provide the $d$ orbital filling factors and quasiparticle weights obtained with on-site interaction $U = 75\,\rm meV$ at different total filling factors $\nu = 1.2, 1.4, 1.6$.
The $d$ orbital filling factors as functions of the displacement field, which are shown in Fig.~\ref{fig:osmp}(a), have ``plateau'' regions in certain $\varepsilon_D$ intervals. The plateaus of different filling factors appear at different displacement field potential values, resulting in a stripe-shaped heavy Fermi liquid region, which is similar to what we obtained in Fig.~\ref{fig:phase_diagrams}(a).
As one would expect from the lack of conduction electron screening, the heavy fermion quasiparticle weight can eventually be diminished to zero with a sufficiently low conduction electron density, which amplifies the contrast between the fate of (de)localization of the two orbitals.
In Fig.~\ref{fig:osmp}(b) we present the displacement field dependent heavy fermion quasiparticle weight with different total filling factors. With smaller total electron densities, the minimum values of $Z_{\rm Mo}$ also get smaller, corresponding to the qualitative trend observed in Fig.~\ref{fig:phase_diagrams}(d). Indeed, $Z_{\rm Mo}$ drops to zero at its minimum point, indicating a transition into the OSMP with a destructed Kondo screening at $\nu = 1.2$.

We then discuss this transition into the OSMP along another axis with fixed displacement field potential $\varepsilon_D = -70\rm\,meV$. 
In Fig.~\ref{fig:osmp}(c), we found that the $d$ orbital filling factor is almost pinned at $\nu_{\rm Mo} = 1$ when the total filling is reduced to $\nu \leq 1.3$. 
It is also noticeable that the heavy fermion quasiparticle weight only drops to zero as the filling factor is further reduced to $\nu \leq 1.2$, as shown in Fig.~\ref{fig:osmp}(d).

We also note that the critical filling factor of the transition into the OSMP strongly depends on the strength of the interaction. For example, as we will discuss in App.~\ref{appsec:osmp-U}, total filling factor $\nu = 1.4$ is in fact sufficiently low for an OSMP, if the on-site interaction is increased to $U = 80\rm \,meV$.

As one would notice in Fig.~\ref{fig:non_interacting_model}(c), since the $d$ orbitals effectively form a triangular lattice in the OSMP, the RKKY interaction might either be frustrated anti-ferromagnetic, or ferromagnetic at very low conduction electron density. The RKKY interaction could lead to either an anti-ferromagnetic ordered state \cite{Kumar2022Gate,Guerci2023Chiral}, a ferromagnetic ordered state \cite{zhao2023emergence}, or a paramagnetic state in the local moment phase, depending on the competition between the frustration strength and the screening strength as mentioned in Fig.~\ref{fig:sketch}(a). However, determining the magnetic order in the OSMP is beyond the scope of the present work.

\section{Discussion}\label{sec:discussion}

\begin{figure}[b!]
    \centering
    \includegraphics[width=\linewidth]{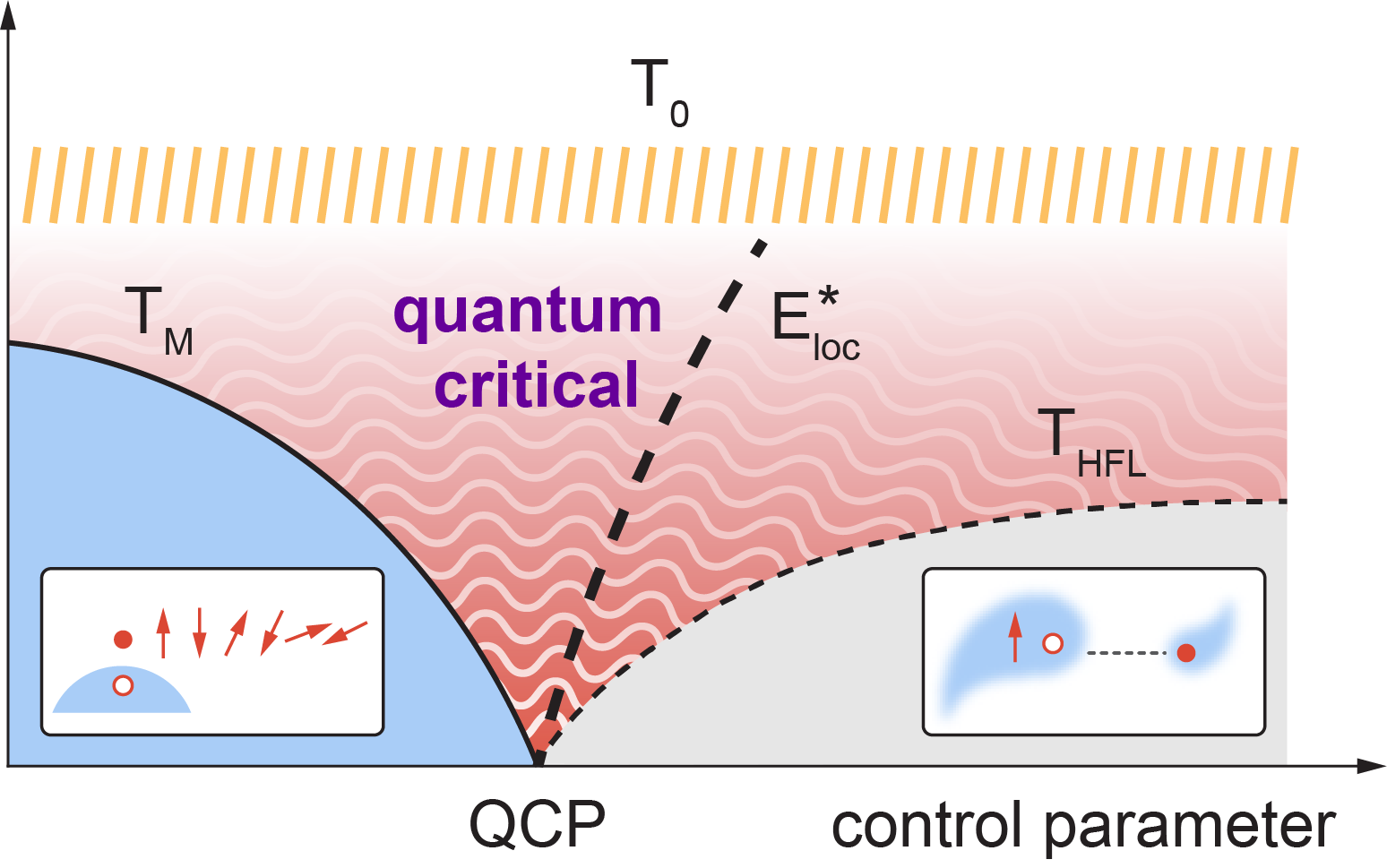}
    \caption{Schematic finite temperature phase diagram 
    near a quantum critical point (QCP)
    to be expected from a dynamical competition between Kondo and RKKY interactions, based on the EDMFT studies of the Kondo lattice \cite{Si2001,Si-JPSJ2014}. 
    The cartoons for the heavy Fermi liquid and Kondo-destroyed phases are adapted from Ref.\,\cite{Prochaska2020}.
    }
    \label{fig:finite-T-sketch}
\end{figure}

We have analyzed the crossover into the heavy Fermi liquid state in the AB-stacked $\rm MoTe_2/WSe_2$ bilayer moir\'e superlattices. 
In addition, we have described the transition from the heavy Fermi liquid state into the  orbital-selective Mott phase that does not contain any long-range order.

To address the quantum critical physics, it is important to study the dynamical interplay between the Kondo effect and RKKY interactions. 
To put the Kondo physics of the AB-stacked $\rm MoTe_2/WSe_2$ in context, we recall that, in the correlation regime captured by the Kondo lattice model, studies based the extended dynamical mean field theory (EDMFT) showed that the RKKY interactions compete against the Kondo effect \cite{Si2001,Colemanetal,senthil2004a}. A new energy scale, $E_{\rm loc}^*$, emerges, characterizing the destruction of the Kondo effect. As illustrated in Fig.\,\ref{fig:finite-T-sketch}, this scale separates two regimes of the phase diagram spanned by temperature and a non-thermal control parameter. To the right of the $E_{\rm loc}^*$ line, the system flows towards a heavy Fermi liquid ground state. The low-energy physics of the heavy quasiparticles and the associated large Fermi surface is characterized by the heavy Fermi liquid temperature scale, $T_{\rm HFL}$. The vanishing of the $T_{\rm HFL}$ scale at the QCP signifies the loss of quasiparticles.
To the left of the $E_{\rm loc}^*$ line, the system flows towards a ground state in which the Kondo singlet is destroyed and the Fermi surface becomes small. Accordingly, the $E_{\rm loc}^*$ line characterizes the localization-delocalization of the heavy fermions.
The role of long-range order depends on where the system lies in the global phase diagram. For example, along trajectory ``I" of Fig.\,\ref{fig:sketch}(a), the onset of the magnetic order is concurrent with the destruction of the Kondo effect. This is also illustrated in Fig.\,\ref{fig:finite-T-sketch}. Finally, the vestige of the coherence temperature is characterized by the temperature scale $T_0$ where the initial onset of the Kondo effect takes place as the temperature is lowered. 
These salient properties have been extensively evidenced by experiments in heavy fermion materials, including the dynamical scaling~\cite{Aro95.1,Schroder,Prochaska2020}, Fermi surface crossover across a $T^*$ temperature scale (associated with the $E_{\rm loc}^*$ energy scale) and its extrapolated zero-temperature jump~\cite{paschen2004,Friedemann09,shishido2005}, a linear-in-$T$ relaxation rate that connects with the $T$-linear electrical resistivity~\cite{Ngu21.1,Par08.1,Geg02.1} and related properties \cite{park-nature06,Kne08.1} and, finally, the loss of quasiparticles \cite{Chen2023Shot}.

The analysis presented here indicates that the AB-stacked hetero-bilayer TMDC opens up a new correlation regime of Kondo destruction. Our findings raises the theoretical question about the  dynamical competition between the Kondo/hybridization and RKKY interactions in this correlation regime, which is left for future work. 
Our results also motivate new opportunities for experimental studies, namely to explore the dynamical signatures of Kondo destruction outlined above in the AB-stacked hetero-bilayer TMDC and related moir\'e structures.

\section{Summary}\label{sec:summary}

In this work, we studied an effective model that can well describe the Kondo physics in the AB-stacked $\rm MoTe_2/WSe_2$ hetero-bilayer system. 
This effective model can be expressed as an extended Hubbard Hamiltonian with $d$ and $c$ orbitals, which correspond to moir\'e bands from the $\rm MoTe_2$ and $\rm WSe_2$ layers, respectively.
The bandwidths of the two moir\'e bands differ by a factor of $\sim 2$, which could lead to orbital-selective correlation effects.
The hybridization between the two orbitals is also taken into consideration. 
The Hubbard Hamiltonian can describe the charge transfer between orbitals with strong displacement fields, which is not captured by the Kondo lattice Hamiltonian.
Interaction in the $c$ orbitals and the bandwidth associated with the $d$ orbitals are also usually not addressed in previous studies.
Hence, we believe that the two-orbital extended Hubbard Hamiltonian is the most suitable model for understanding the Kondo-driven phenomena in this moir\'e structure without unnecessary complexity.

We then use a saddle-point approach to solve this effective interacting Hamiltonian.
Using the total electron filling factor and the displacement field as tunable parameters, we identify the crossover into a heavy Fermi liquid in the phase diagram, which resembles the experimentally observed results \cite{zhao2022gate}. 
With the $d$ orbital bandwidth considered, our numerical study can also explain the heavy electron mass $m_{\rm Mo}\approx 10m_e$ and the high coherence temperature $T_{\rm coh}\approx 20\sim 40\,\rm K$, which have been observed in transport measurements, providing a consistent interpretation.
These results indicate that the Kondo effect in this system is in an extended correlation regime that goes beyond the canonical Kondo-lattice description.

We have also explored the consequence of doping the $d$ orbital away from half filling. This doping, $\delta=\nu_d - 1$, serves as a new axis of the heavy fermion phase diagram.
The numerical results have shown that the heavy Fermi liquid states with a very small $Z_{\rm Mo}$ can still exist in a narrow window when $\delta = \nu_d - 1 \neq 0$. When the $d$ orbital is doped noticeably away from half filling, a small displacement field change can lead to a drastic change of its value, and the heavy fermion quasiparticle weight as well as the coherence temperature will be significantly increased. These phenomena have also been observed in the experiments as well. 

In addition, our approach captures the orbital-selective Mott phase, in which the $d$ electrons are localized while the $c$ electrons remain itinerant.
Our saddle-point analysis shows a vanishing heavy fermion quasiparticle weight at sufficiently low electron density.
This is also qualitatively in agreement with the recent experimental observation \cite{zhao2023emergence}.

Our results set the stage to address the amplified quantum fluctuations that the Kondo effect may produce in these moir\'e structures and the new correlation regimes that they open up for the Kondo-destruction quantum criticality.
As such, the TMDC moir\'e structures provides a new setting to explore the salient properties associated with the amplified quantum fluctuations, such as the dynamical $\hbar \omega /k_{\rm B} T$ scaling, the linear-in-$T$ relaxation rate, and loss of quasiparticles.

\begin{acknowledgments}
We thank Jennifer Cano, Kin Fai Mak, Jie Shan, and Wenjin Zhao for helpful discussions.
This work has primarily been supported by the U.S. DOE, BES, under Award No. DE-SC0018197 (model construction, F.X. and L.C.), by the Air Force Office of Scientific Research under Grant No. FA9550-21-1-0356 (conceptualization and model calculation, F.X., L.C. and Q.S.), and by the Robert A. Welch Foundation Grant No. C-1411 and the Vannevar Bush Faculty Fellowship No. ONR-VB N00014-23-1-2870 (Q.S.). 
The majority of the computational calculations have been performed on the Shared University Grid at Rice funded by NSF under Grant No.~EIA-0216467, a partnership between Rice University, Sun Microsystems, and Sigma Solutions, Inc., the Big-Data Private-Cloud Research Cyberinfrastructure MRI-award funded by NSF under Grant No. CNS-1338099, and the Extreme Science and Engineering Discovery Environment (XSEDE) by NSF under Grant No. DMR170109. 
We acknowledge the hospitality of the Kavli Institute for Theoretical Physics, supported in part by the National Science Foundation under Grant No. NSF PHY1748958, during the program ``A Quantum Universe in a Crystal: Symmetry and Topology across the Correlation Spectrum.''
Q.S. acknowledges the hospitality of the Aspen Center for Physics, which is supported by NSF grant No. PHY-2210452.

\end{acknowledgments}

\bibliography{tmd.bib}

\onecolumngrid
\tableofcontents
\appendix

\section{Model}\label{appsec:model}

In this appendix, we provide a detailed discussion about the non-interacting Hamiltonian and its Wannier functions in App.~\ref{appsubsec:wannier}. A brief review of the continuum model is discussed, and Wannier functions for the low-energy moir\'e bands can be constructed by analyzing their symmetry eigenvalues. The tight-binding model introduced in Sec.~\ref{sec:tightbinding} is based on the Wannier functions obtained from the continuum model.

The Coulomb interaction matrix elements discussed in App.~\ref{appsubsec:interaction} are also based on the Wannier functions obtained from the continuum model. The effective interacting Hamiltonian in Eq.~(\ref{eq:Hint}) is guided by these matrix elements as well.

\subsection{Wannier functions of the non-interacting Hamiltonian}\label{appsubsec:wannier}
We first review the continuum model of twisted bilayer TMDC materials. 
In this system, the moir\'e superlattice is originated from the lattice mismatch between the $\rm MoTe_2$ and $\rm WSe_2$ layers. Both layers are hexagonal lattices with lattice constants $a_{\rm Mo} \approx 3.575 \textrm{\AA}$ and $a_{\rm W} \approx 3.32\textrm{\AA}$ \cite{Zhang2021Spin}. Therefore, the lattice constant of the moir\'e unit cell is $a_{\rm M} = {a_{\rm Mo} a_{\rm W}}/({a_{\rm Mo} - a_{\rm W}}) \approx 46.55 \textrm{\AA}$, which is much larger than the lattice constant of each individual layer. As a consequence, the reciprocal lattice of the moir\'e lattice is much smaller than the Brillouin zones of the single layers, which can be seen in Fig.~\ref{fig:non_interacting_model}(b). In this paper, we use $\mathbf{a}_1 = (0, a_M), \mathbf{a}_2 = (a_M/2, \sqrt{3} a_M/2)$ to represent the moir\'e lattice basis vectors, and use $\mathbf{b}_1 = (2\pi/a_M, -2\pi/\sqrt{3}a_M)$, $\mathbf{b}_2 = (0, 4\pi/\sqrt{3}a_M)$ to represent the corresponding reciprocal basis vectors.

Similar to the Bistritzer-MacDonald model for twisted bilayer graphene, we can also write down the Hamiltonian for AB-stacked bilayer $\rm MoTe_2/WSe_2$ in a {\it single valley} as follows:
\begin{equation}
    h = \left(
    \begin{array}{cc}
        h_{\hat{d}}(-i\bm{\partial}) + V_{\hat{d}}(\mathbf{r}) - \frac{\varepsilon_D}{2} & W(\mathbf{r})\\
        W^*(\mathbf{r}) & h_{\hat{c}}(-i\bm{\partial}) + V_{\hat{c}}(\mathbf{r}) + \frac{\varepsilon_D}{2}
    \end{array}
    \right)\,,
\end{equation}
in which $h_{\hat{d}}$, $h_{\hat{c}}$ are the quadratic dispersion Hamiltonian of the individual layers:
\begin{equation}
    h_{\hat{d},\hat{c}}(-i\bm{\partial}) = \frac{1}{2m_{\hat{d},\hat{c}}}\bm{\partial}^2\,.
\end{equation}
$V_{\hat{d}}(\mathbf{r})$ and $V_{\hat{c}}(\mathbf{r})$ are the moir\'e potential received from the other layer, and $W(\mathbf{r})$ is the interlayer tunneling strength, which has the same period as the moir\'e superlattice. As mentioned in Ref.~\cite{Zhang2021Spin}, the interlayer tunneling is strongly suppressed, since it is a spin-flipping process. Therefore, it is reasonable to build the Wannier functions for each layer individually.

Previous density functional theory studies have shown that the effective band mass of the $\rm MoTe_2$ layer is $m_{\hat{d}} = 0.65 m_e$, and the mass of the $\rm WSe_2$ layer is $m_{\hat{c}} = 0.35m_e$, in which $m_e = 9\times 10^{-31}\,\rm kg$ is the bare electron mass \cite{Zhang2021Spin}. The intralayer moir\'e potential has the following form:
\begin{equation}
    V_{\hat{d}, \hat{c}}(\mathbf{r}) = 2v_{\hat{d}, \hat{c}}\sum_{i=1}^3\cos(\mathbf{g}_i\cdot \mathbf{r} + \phi_{\hat{d}, \hat{c}})\,,
\end{equation}
where $\mathbf{g}_i = C^{i -1 }_3 \mathbf{g}_1$ and $\mathbf{g}_1 = \mathbf{b}_1$ are the three smallest moir\'e reciprocal vectors along different directions. The eigenstates of the single layer Hamiltonian in the valley $\tau$ are Bloch states with the following form:
\begin{equation}
    \psi_{\vk,\tau\alpha}(\mathbf{r}) = \frac{1}{\sqrt{\Omega_{\rm tot}}}\sum_\mathbf{Q} u_{\mathbf{Q}, \tau\alpha}(\vk) e^{i(\vk - \mathbf{Q})\cdot \mathbf{r}}\,.
\end{equation}
Here $\mathbf{Q}$ stands for moir\'e lattice reciprocal vectors, and $\alpha = \hat{c}, \hat{d}$ stands for the two types of fermions. The wavefunctions in the opposite valley $-\tau$ can be obtained via a time reversal transformation.

\begin{figure}
    \centering
    \includegraphics[width=0.75\linewidth]{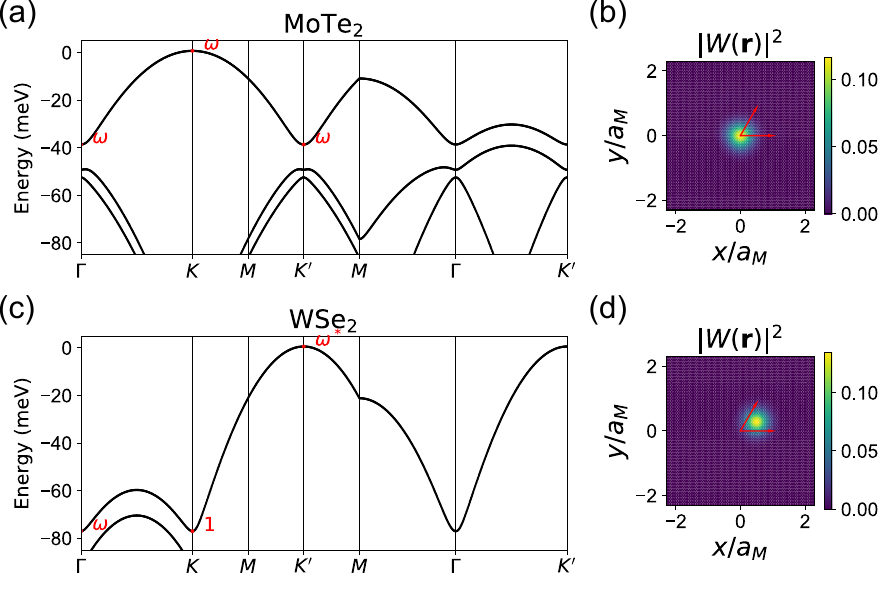}
    \caption{(a) The single valley band structure of the $\rm MoTe_2$ layer without interlayer tunneling considered. The band structure in the opposite valley can be obtained by simply applying time reversal transformation to the band structure in this valley. The (spinless) $C_{3z}$ eigenvalues at high symmetry points $\Gamma$, $K$ and $K'$ are labeled, in which $\omega = e^{i\frac{2\pi}{3}}$. (b) The Wannier orbital of the highest band (the lowest hole band) in the $\rm MoTe_2$ layer. (c) The single valley band structure of the $\rm WSe_2$ layer without interlayer tunneling.  (d) The Wannier orbital of the highest band in the $\rm WSe_2$ layer. These wavefunctions are obtained via Wannier90 \cite{Pizzi2020wannier}, and the red arrows represent the basis vectors of the Bravais moir\'e lattice $\mathbf{a}_1$ and $\mathbf{a}_2$.}
    \label{fig:wannier}
\end{figure}

In the $\rm MoTe_2$ layer, the values of these potential parameters are $v_{\hat{d}} = 4.1{\,\rm meV}, \phi_{\hat{d}} = 14^\circ$ \cite{Zhang2021Spin}. Solving the kinetic Hamiltonian of this layer yields the band structure shown in Fig.~\ref{fig:wannier}(a). The $C_{3z}$ eigenvalues of the three high symmetry points are also labeled. By comparing these eigenvalues with the little group irreps in Table~\ref{tab:little-group} and the EBRs in Table~\ref{tab:ebr}, we can find that this moir\'e band indeed corresponds to a local orbital on Wyckoff position $1a$ with site symmetry group representation ${}^2E$. Using Wannier90 \cite{Pizzi2020wannier} program with a trial wavefunction at this Wyckoff position and the angular momentum of ${}^2E$, we are able to find the proper gauge choice of the Bloch wavefunction $u_{\mathbf{Q},\tau\alpha}(\vk)$ which corresponds to the maximally localized Wannier function of the moir\'e band \cite{marzari_maximally_2012}.
The real space distribution of its orbital has been shown in Fig.~\ref{fig:wannier}(b). Clearly that this orbital is highly concentrated within the size of a moir\'e unit cell.

It has been mentioned in Refs.~\cite{Zhang2021Spin,Pan2022Topological,Rademaker2022Spin} that the detail of the intralayer moir\'e potential of the $\rm WSe_2$ layer $V_{\hat{c}}(\mathbf{r})$ is not crucial to the low energy physics. However, it still controls the little group representation at high symmetry points, since the gap opening around $\Gamma$ and $K$ is very sensitive to the phase $\phi_{\hat{c}}$. Fortunately, Ref.~\cite{Zhang2021Spin} provided the $C_{3z}$ eigenvalues of the $\rm WSe_2$ band at these high symmetry points, which indicates that $60^\circ \leq\phi_{\hat{c}} \leq 180^\circ$. Therefore, we use $v_{\hat{c}} \approx 5\rm \,meV$ and $\phi_{\hat{c}} \approx 120^\circ$ as estimated values for the intralayer potential. The corresponding single layer band structure of $\rm WSe_2$ layer can be found in Fig.~\ref{fig:wannier}(c). $C_{3z}$ eigenvalues at high symmetry points also agree with the ${}^2E$ elementary band representation at Wyckoff position $1c$. We also show the Wannier orbital obtained by Wannier90 for the $\rm WSe_2$ layer in Fig.~\ref{fig:wannier}(d). This explains why the two orbitals formed in two layers give rise to a hexagonal lattice, as depicted in Fig.~\ref{fig:non_interacting_model}(c) in the main text.

\begin{table}
\begin{center}
\begin{tabular}{c|ccc|ccc|ccc}
    \hline
    & $\Gamma_1$ & $\Gamma_2$ & $\Gamma_3$ & $K_1$ & $K_2$ & $K_3$ & $K'_1$ & $K'_2$ & $K'_3$\\
    \hline
    $E$ & 1 & 1 & 1 & 1 & 1 & 1 & 1 & 1 & 1\\
    $C_{3z}$ & 1 & $\omega$ & $\omega^*$ & 1 & $\omega$ & $\omega^*$ & 1 & $\omega^*$ & $\omega$\\
    \hline
\end{tabular}
\caption{The character table of irreps at high symmetry points of space group 143.}
\label{tab:little-group}
\end{center}
\end{table}

\begin{table}
    \begin{center}
    \begin{tabular}{c|c|c|c|c|c|c|c|c|c}
        \hline
        Wyckoff positions & \multicolumn{3}{c|}{$1a$} & \multicolumn{3}{c|}{$1b$} & \multicolumn{3}{c}{$1c$}\\
        \hline
        EBR & $A_1\uparrow G$ & ${}^1E\uparrow G$ & ${}^2E\uparrow G$ & $A_1\uparrow G$ & ${}^1E\uparrow G$ & ${}^2E\uparrow G$ & $A_1\uparrow G$ & ${}^1E\uparrow G$ & ${}^2E\uparrow G$ \\
        \hline
        $\Gamma$ & $\Gamma_1$ & $\Gamma_3$ & $\Gamma_2$ & $\Gamma_1$ & $\Gamma_3$ & $\Gamma_2$ & $\Gamma_1$ & $\Gamma_3$ & $\Gamma_2$ \\
        \hline
        $K$ & $K_1$ & $K_3$ & $K_2$ & $K_2$ & $K_1$ & $K_3$ & $K_3$ & $K_2$ & $K_1$ \\
        \hline
        $K'$ & $K'_1$ & $K'_2$ & $K'_3$ & $K'_2$ & $K'_3$ & $K'_1$ & $K'_3$ & $K'_1$ & $K'_2$\\
        \hline
    \end{tabular}
    \caption{The elementary band representation of space group 143 of the three maximum Wyckoff positions.}
    \label{tab:ebr}
\end{center}
\end{table}

\subsection{The projection of Coulomb interaction}\label{appsubsec:interaction}
In this subsection, we use some realistic parameters to estimate the interaction strength of our effective four-band model. 
With the Wannier wave functions obtained in the previous subsection, we are able to project the screened Coulomb interaction into these low-energy orbitals. The projected interactions can be written as:
\begin{equation}
    H_I = \frac{1}{2\Omega_{\rm tot}} \sum_{\vk\vk'\vq\in{\rm MBZ}}\sum_{\tau\tau'=\pm}\sum_{\alpha\alpha'=\hat{c},\hat{d}}U^{\tau\tau'}_{\alpha\alpha'}(\vq; \vk,\vk')\alpha^\dagger_{\vk + \vq, \tau}\alpha_{\vk, \tau}\alpha'^\dagger_{\vk' - \vq, \tau'}\alpha'_{\vk', \tau'}\,.
\end{equation}
Here we still use $\tau,\tau'=\pm$ to represent the two valleys $\mathcal{K}$ and $\mathcal{K}'$, and $\alpha, \alpha' = \hat{c},\hat{d}$ to represent the two flavors ($\rm MoTe_2$ layer and $\rm WSe_2$ layer) of fermions. The interaction matrix elements $U^{\tau\tau'}_{\alpha\alpha'}(\vq;\vk,\vk')$ can be written in the following form:
\begin{equation}
    U^{\tau\tau'}_{\alpha\alpha'}(\vq;\vk,\vk') = \sum_{\mathbf{G}}V(\vq + \mathbf{G})M^{\tau}_\alpha(\vk, \vq + \mathbf{G})M^{\tau'}_{\alpha'}(\vk', -\vq - \mathbf{G})\,,
\end{equation}
in which $\mathbf{G}$ are reciprocal vectors of the moir\'e lattice. The Fourier transformed screened Coulomb potential has the form $V(\vq) = (\xi e^2/4\varepsilon_0\varepsilon)\tanh(\xi q/2)/(\xi q/2)$, where we use $\xi \approx 10\, \rm nm$ as the distance between the gates, and $\varepsilon \approx 10$ as the dielectric constant of the substrate. 
The form factors $M^\tau_\alpha(\vk, \vq + \mathbf{G})$ can be expressed by the inner products of the non-interacting Bloch wave functions:
\begin{equation}
    M^\tau_\alpha(\vk, \vq + \mathbf{G}) = \sum_{\mathbf{Q
    }} u^*_{\mathbf{Q} + \mathbf{G}, \tau \alpha}(\vk + \vq) u_{\mathbf{Q}, \tau \alpha}(\vk)\,,
\end{equation}
where we have to use the gauge choice of $u_{\mathbf{Q},\tau\alpha}(\vk)$, which corresponds to the maximally localized Wannier states.
Performing a discrete Fourier transformation into the Wannier function basis, this projected interacting Hamiltonian can also be written as:
\begin{equation}
    H_I = \frac{1}{2}\sum_{\mathbf{R}_0}\sum_{\mathbf{R}\mathbf{D}\mathbf{D}'}\sum_{\tau\tau', \alpha\alpha'}\tilde{U}^{\tau\tau'}_{\alpha\alpha'}(\mathbf{R};\mathbf{D},\mathbf{D}')\alpha^\dagger_{\mathbf{R} + \mathbf{D}+\mathbf{R}_0, \eta, l}\alpha_{\mathbf{R} + \mathbf{R}_0, \eta, l}\alpha'^\dagger_{\mathbf{D}' + \mathbf{R}_0, \eta',l'}\alpha'_{\mathbf{R}_0, \eta', l'}\,.
\end{equation}
Obviously, the real space interaction elements are given by the following expression:
\begin{equation}
    \tilde{U}^{\tau\tau'}_{\alpha\alpha'}(\mathbf{R};\mathbf{D},\mathbf{D}') = \frac{1}{\Omega_c N_M^3}\sum_{\vk\vk'\vq\in\mathrm{MBZ}}U^{\tau\tau'}_{\alpha\alpha'}(\vq;\vk,\vk')e^{i\vq\cdot(\mathbf{R} + \mathbf{D} - \mathbf{D}')}e^{i\vk\cdot \mathbf{D}}e^{i\vk'\cdot\mathbf{D}'}\,.
\end{equation}
Usually, two types of these projected interaction terms are considered when building an interacting lattice model. The first type of terms belongs to the direct channel with $\mathbf{D} = \mathbf{D}' = 0$:
\begin{equation}
    \tilde{V}^{\tau\tau'}_{\alpha\alpha'}(\mathbf{R}) = \tilde{U}^{\tau\tau'}_{\alpha\alpha'}(\mathbf{R};\mathbf{D}=0,\mathbf{D}'=0)\,,
\end{equation}
and the other type of terms belongs to the exchange channel with $\mathbf{D} = -\mathbf{D}' = -\mathbf{R} \neq 0$:
\begin{equation}
    \tilde{J}^{\tau\tau'}_{\alpha}(\mathbf{R}) = -\tilde{U}^{\tau\tau'}_{\alpha\alpha}(\mathbf{R};\mathbf{D}=-\mathbf{R},\mathbf{D}'=\mathbf{R})\,.
\end{equation}
Using the Coulomb potential $V(\vq)$ and the Bloch state wave functions we obtained in the previous subsection, we numerically evaluated the values of some direct interaction terms, which can be found in Table \ref{tab:v-values}. The on-site interactions $U$ of the two layers $\tilde{V}_{\hat{c}\hat{c}}(0)$ and $\tilde{V}_{\hat{d}\hat{d}}(0)$ are close but different. As a simplification, we use the same value for both layers in the actual 
calculation. The nearest-neighbor interaction $V = \tilde{V}_{\hat{c}\hat{d}}(0)$ approximately satisfies $V \approx U/2$, and the next-nearest-neighbor interaction $V' =  \tilde{V}(\mathbf{a}_{1,2})$ is also around $V/2$. Hence, choosing $U = 70\rm\,meV$, $V = U/2$ and $V' = U/4$ are reasonable estimations for these parameters. Since the screened Coulomb potential can change with the dielectric constant $\varepsilon$ and the gate distance $\xi$, it is important to stress that these on-site, NN, and NNN interaction values should be regarded as reasonable estimations, and they can have different values in realistic materials.

On the other hand, the largest exchange term we obtained is only around $1.05\rm\, meV$, thus we ignored all the exchange interactions.

\begin{table}[htbp]
    \centering
    \begin{tabular}{c|c|c|c|c|c}
    \hline \xrowht{10pt}
      & $\tilde{V}_{\hat{d}\hat{d}}(0)$  & $\tilde{V}_{\hat{c}\hat{c}}(0)$ & $\tilde{V}_{\hat{c}\hat{d}}(0)$ & $\tilde{V}_{\hat{d}\hat{d}}(\mathbf{a}_1)$ & $\tilde{V}_{\hat{c}\hat{c}}(\mathbf{a}_1)$\\
   \hline \xrowht{10pt}
      Value (meV) & 70.14 & 64.08 & 37.76 & 16.52 & 16.38\\
    \hline
    \end{tabular}
    \caption{The numerical values of the largest direct interaction terms.}
    \label{tab:v-values}
\end{table}

\section{Method}\label{appsec:method}

To take into account the interaction effect, we utilize the $U(1)$-slave spin (SS) method~\cite{Yu2012U1slave} to solve the effective Hamiltonian described in Eqs.~(\ref{eq:h0}) and (\ref{eq:Hint}). We introduce the slave spin representation for a fermion operator as a product of an auxiliary bosonic operator $o^\dagger$ and an auxiliary fermionic operator $f^\dagger$:
\begin{equation}
    \alpha^{\dagger}_{i\alpha\tau} = o_{i\alpha\tau}^{\dagger} f^{\dagger}_{i\alpha\tau}\,.
\end{equation}
Here $\alpha^\dagger$ enumerates the creation operators of $c$ orbitals $c^\dagger$ or $d$ orbitals $d^\dagger$. The auxiliary bosonic field $o^{\dagger}_{i\alpha\tau} = P^{+}_{i\alpha\tau} S^{+}_{i\alpha\tau}P^{-}_{i\alpha\tau}$ is represented by the product of a spin operator $S^+_{i\alpha\tau}$ and projection operators $P^{\pm}_{i\alpha\tau} = \frac{1}{\sqrt{1/2\pm S_{i\alpha\tau}^z}}$, which is suitable for a system away from half filling. By introducing these auxiliary operators, the physical Fock states can be mapped to the enlarged Hilbert space of the slave fermion and slave spin operators:
\begin{equation}
    \begin{aligned}
        |n_{i\alpha\tau}=0 \rangle \leftrightarrow    |n_{i\alpha\tau}^{f}=0 \rangle | S^{z}_{i\alpha\tau} = \downarrow \rangle\,,  \\
        |n_{i\alpha\tau}=1 \rangle \leftrightarrow    |n_{i\alpha\tau}^{f}=1 \rangle | S^{z}_{i\alpha\tau} = \uparrow \rangle\,,  \\
        \varnothing \leftrightarrow    |n_{i\alpha\tau}^{f}=0 \rangle | S^{z}_{i\alpha\tau} = \uparrow \rangle\,,  \\
        \varnothing  \leftrightarrow    |n_{i\alpha\tau}^{f}=1 \rangle | S^{z}_{i\alpha\tau} = \downarrow \rangle\,.  \\
    \end{aligned}
\end{equation}
We notice that the introduction of the auxiliary operators expands the Hilbert space. Therefore, the following constraint is required to project out the unphysical states:
\begin{equation}\label{eq:cont}
    S_{i\alpha\tau}^{z}+\frac{1}{2} = n_{i\alpha\tau}^{f}\,.
\end{equation}
At the mean-field level, we treat this constraint on average by introducing the Lagrange multipliers $\lambda_{i\alpha\tau}$, such that $\langle S^z_{i\alpha\tau}\rangle + \frac{1}{2} =  n^{f}_{i\alpha\tau} $. Note that in the physical Hilbert space, $n^{f}_{i\alpha\tau} = n_{i\alpha\tau}$. Using the slave fermion and spin operators, the kinetic Hamiltonian can be written as:
\begin{equation}
    \begin{aligned}
        H_0 &= \sum_{\langle\langle i, j\rangle\rangle,\alpha\tau} t_{\alpha}e^{\tau i\phi_{ij}^{\alpha}}  f_{i\alpha\tau}^{\dagger}o^{\dagger}_{i\alpha\tau}  o_{j\alpha\tau}f_{j\alpha\tau} +\sum_{\langle i, j\rangle,\tau } \left( t_{cd} f^{\dagger}_{ic\tau} o^{\dagger}_{ic\tau} o_{jd\tau}f_{jd\tau} +{\rm h.c.} \right) + \frac{\varepsilon_{D}}{2} \sum_{i\tau} \left( f_{ic\tau}^{\dagger}f_{ic\tau} - f_{id\tau}^{\dagger}f_{id\tau}  \right).
    \end{aligned}
\end{equation}
We can also decompose the product of auxiliary fermion and slave spin operators by the following equation:
\begin{equation}
    f_{i\alpha\tau}^{\dagger} o_{i\alpha\tau}^{\dagger}o_{j\alpha'\tau'} f_{j\alpha'\tau'} = f_{i\alpha\tau}^{\dagger}f_{j\alpha'\tau'} \langle o_{i\alpha\tau}^{\dagger}o_{j\alpha'\tau'}\rangle + \langle f_{i\alpha\tau}^{\dagger}f_{j\alpha'\tau'} \rangle o_{i\alpha\tau}^{\dagger}o_{j\alpha'\tau'} +{\rm h.c.}\,.
\end{equation}
Furthermore, products of auxiliary bosonic operators can be decoupled via the single-site approximation $o_{i\alpha\tau}^{\dagger}o_{j\alpha'\tau'} = \langle o_{i\alpha\tau}^{\dagger}\rangle o_{j\alpha'\tau'} +  o_{i\alpha\tau}^{\dagger}\langle o_{j\alpha'\tau'}\rangle - \langle o_{i\alpha\tau}^{\dagger}\rangle \langle o_{j\alpha'\tau'}\rangle$. By assuming that the bosonic operator expectation values  are translation invariant $\langle o_{i\alpha\tau} \rangle = \langle o_{j\alpha\tau} \rangle = \langle o_{\alpha\tau}\rangle$, the two mean-field Hamiltonians for slave fermion and slave spin are given by:
\begin{equation}
    \begin{aligned}
        H^{f} &= \sum_{\langle \langle ij\rangle \rangle,\alpha\tau} t_{\alpha} e^{\tau i\phi_{ij}^{\alpha}} \langle o_{\alpha\tau}^{\dagger} \rangle \langle o_{\alpha\tau} \rangle f_{i\alpha\tau}^{\dagger}f_{i\alpha\tau} + \sum_{\langle i,j \rangle, \tau} \left(t_{cd} \langle o_{c\tau}^{\dagger} \rangle \langle o_{d\tau} \rangle f_{ic\tau}^{\dagger} f_{jd\tau} + {\rm h.c.}  \right) +  \frac{\varepsilon_{D}}{2} \sum_{i\tau} \left( f_{ic\tau}^{\dagger}f_{ic\tau} - f_{id\tau}^{\dagger}f_{id\tau} \right)\,, \\
        H^S &=  \sum_{i\alpha\tau}\sum_{\vk} \left( \epsilon^\tau_{\alpha}(\vk) \langle f_{\vk \alpha \tau}^{\dagger}f_{\vk \alpha \tau}\rangle \langle o_{\alpha\tau}\rangle o_{i\alpha\tau}^{\dagger} + {\rm h.c.} \right) + \sum_{i\tau} \sum_{\vk} \left( \epsilon_{cd} (\vk)\langle f_{\vk c\tau}^{\dagger} f_{\vk d \tau}\rangle  \left(\langle o^{\dagger}_{c\tau}\rangle o_{id\tau} + o^{\dagger}_{ic\tau}\langle o_{d\tau} \rangle \right) + {\rm h.c.}  \right) + H^S_{I},
    \end{aligned}
\end{equation}
where $\epsilon^\tau_{\alpha}(\vk)$ and $\epsilon_{cd}(\vk)$ are the Fourier transformation of the intra and inter-orbital hopping terms in the momentum space, respectively. 
We further perform the Taylor expansion of the projection operators $P^\pm_{i\alpha\tau}$ and bosonic operators $o^\dagger_{i\alpha\tau}$ by considering $\delta S^z_{i\alpha\tau} = S^z_{i\alpha\tau} - \langle S^{z}_{\alpha\tau}\rangle$ as a small parameter:
\begin{equation}
\begin{aligned}
    P^{\pm}_{i\alpha\tau} &\approx \frac{1}{\sqrt{\frac{1}{2} \pm \langle S^{z}_{\alpha\tau}\rangle}} \left( 1 \mp \frac{\delta S^z_{i\alpha\tau}}{\sqrt{ \frac{1}{2}\pm \langle S^{z}_{\alpha\tau}\rangle}}\right)\,,\\
    o_{i\alpha\tau}^{\dagger} & \approx \langle P^{+}_{\alpha\tau} \rangle S^{+}_{i\alpha\tau} \langle P_{\alpha\tau}^{-}\rangle + \langle P^{+}_{\alpha\tau} \rangle \langle S^{+}_{i\alpha\tau} \rangle \langle P_{\alpha\tau}^{-}\rangle \frac{1}{2} (S^{z}_{i\alpha\tau} - \langle S^{z}_{\alpha\tau} \rangle) \left( \frac{-1}{n_{\alpha\tau}} + \frac{1}{1-n_{\alpha\tau}} \right)\,,  \\
    &  = O^{\dagger}_{i\alpha\tau} + \langle O^{\dagger}_{\alpha\tau} \rangle \eta_{\alpha\tau} \left[ 2S^{z}_{i\alpha\tau} - (2n_{\alpha\tau} - 1) \right]\,, 
\end{aligned}
\end{equation}
where $O^{\dagger}_{i\alpha\tau} = \langle P^{+}_{\alpha\tau}\rangle S^{+}_{i\alpha\tau} \langle P^-_{\alpha\tau} \rangle $, $\eta_{\alpha\tau} = \frac{1}{2} \frac{n_{\alpha\tau}-1/2}{(1-n_{\alpha\tau})n_{\alpha\tau}}$, and $\langle o_{\alpha\tau}\rangle = \langle O_{\alpha\tau}\rangle$. Using these expansions, the slave spin Hamiltonian can be written as the following form:
\begin{equation}
\begin{aligned}
    H^{S} &= \sum_{i\alpha\tau} \left( \Tilde{\epsilon}_{\alpha}  \langle O_{\alpha\tau}\rangle \left( O_{i\alpha\tau}^{\dagger} + \langle O^{\dagger}_{\alpha\tau}\rangle \eta_{\alpha\tau}[2S^z_{i\alpha\tau} - (2n_{\alpha\tau} -1)]  \right) +{\rm h.c.} \right) \\
    & + \sum_{i\tau} \big[ \Tilde{\epsilon}_{cd} \langle O_{c\tau}^{\dagger} \rangle \left( O_{id\tau} +\langle O_{d\tau} \rangle \eta_{d\tau} [2S^{z}_{id\tau} - (2n_{d\tau} - 1)] \right) \\
    & +  \Tilde{\epsilon}_{cd}\left( O_{ic\tau}^{\dagger}  + \langle O_{c\tau}^{\dagger}\rangle \eta_{c\tau}[2S^z_{ic\tau} - (2n_{c\tau} - 1)]  \right) \langle O_{d\tau} \rangle + {\rm h.c.} \big] + H^S_{I}\,, \\
\end{aligned}
\end{equation}
in which $\Tilde{\epsilon}_{\alpha} = \sum_{\vk} \epsilon_{\alpha}(\vk)\langle f^{\dagger}_{\vk\alpha\tau} f_{\vk\alpha\tau}\rangle $ and $\Tilde{\epsilon}_{cd} = \sum_{\vk} \epsilon_{cd}(\vk) \langle f_{\vk c\tau}^{\dagger}f_{\vk d\tau}\rangle$. We use the relationship $2S^z_{i\alpha\tau} = f_{i\alpha\tau}^{\dagger}f_{i\alpha\tau} - 1$ to replace terms proportional to $S^z_{i\alpha\tau}$ and move them from $H^{S}$ to $H^{f}$ and combine the constraint Eq.~(\ref{eq:cont}). This procedure could completely decouple the slave fermion and slave spin Hamiltonians:
\begin{equation}
    \begin{aligned}
           H^{f} &= \sum_{\langle\langle i, j\rangle\rangle,\tau} \left( \langle O_{c\tau }^{\dagger} \rangle \langle O_{c\tau } \rangle t_{c} e^{i\tau\phi_{ij}^c} f^{\dagger}_{ic\sigma}f_{jc\tau} + {\rm h.c.}\right) 
            + \sum_{\langle\langle i, j\rangle\rangle,\tau}  \left( \langle O_{d\tau }^{\dagger} \rangle \langle O_{d\tau } \rangle t_{d} e^{i\tau\phi_{ij}^d} f^{\dagger}_{id\sigma}f_{jd\tau} + h.c.\right) \\
            & + \sum_{\langle i,j \rangle\tau} \left( \langle O_{c\tau }^{\dagger} \rangle \langle O_{d\tau } \rangle t_{cd}  f_{ic\tau}^{\dagger} f_{jd\tau} + {\rm h.c.}\right) \\
            & + \sum_{i\tau} \left( \frac{\epsilon_D}{2} - \mu -\lambda_{c\tau} + \lambda_{c\tau}^{0} \right) f_{ic\tau}^{\dagger} f_{ic\tau} + \left( \frac{-\epsilon_D}{2} - \mu -\lambda_{d\tau} + \lambda_{d\tau}^{0} \right) f_{id\tau}^{\dagger} f_{id\tau}\,, \\ 
             H^{S} &= \sum_{i\tau} \left[ \Tilde{\epsilon}_{c} \left( \langle O_{c\tau}\rangle O_{ic\tau}^{\dagger} + {\rm h.c.} \right) + \Tilde{\epsilon}_{d} \left( \langle O_{d\tau} \rangle O_{id\tau}^{\dagger} + {\rm h.c.} \right) + \Tilde{\epsilon}_{cd} \left( \langle O_{c\tau}\rangle O_{id\tau}^{\dagger} + {\rm h.c.} \right)\right] \\
    & + \sum_{i\tau\alpha} \lambda_{\alpha\tau} \left( S^{z}_{i\alpha\tau}+\frac{1}{2}\right) +H^{S}_{I}
    \end{aligned}
\end{equation}
where we further introduce the chemical potential $\mu$ to control the filling factor and $\lambda_{\alpha\tau}^0=2 (\bar{\epsilon}_{\alpha\alpha} + \bar{\epsilon}_{\alpha\bar{\alpha} } )\eta_{\alpha\tau}$, with $\bar{\epsilon}_{\alpha\alpha'} =\left(\Tilde{\epsilon}_{\alpha\alpha'}\langle O_{\alpha\tau}\rangle \langle O_{\alpha'\tau}^{\dagger}\rangle + {\rm c.c.}\right)$. The interaction part of the slave spin Hamiltonian $H^S_I$ contains the on-site terms $U$, as well as the NN and NNN terms treated on the mean-field level:
\begin{equation}
    H^{S}_{I} = \sum_i\left(\frac{U}{2} \sum_{\alpha} \left( \sum_{\tau} S^z_{i\alpha\tau} \right)^2 + V\sum_{\tau\tau'} \left( 3 S_{ic\tau}^z \langle S^z_{d\tau'} \rangle + 3 \langle S^{z}_{c\tau}\rangle S^z_{id\tau'}\right) +V'\sum_{\alpha\tau\tau'}  6 S^{z}_{i\alpha\tau} \langle S^z_{\alpha\tau'}\rangle \right) \,.
\end{equation}
The numbers $3$ and $6$ come from the coordination numbers of NN and NNN interactions. Due to the mean field treatment of these interaction terms, different unit cells in the slave spin Hamiltonian $H^S$ become completely decoupled from each other:
\begin{equation}
    H^S = \sum_i H^S_i\,.
\end{equation}
Thus the mean field equations can be solved self-consistently with effectively only one unit cell in the slave spin Hamiltonian, by assuming $\langle O_{\alpha\tau} \rangle$, $\lambda_{\alpha\tau}$ and $\mu$ as variational parameters. The quasiparticle weight associated with orbital $\alpha$ is described by $Z_{\alpha\tau} = \langle O_{\alpha\tau}\rangle \langle O^{\dagger}_{\alpha\tau} \rangle = |z_{\alpha\tau}|^2$ if the self-consistently solved value of $z_{\alpha\tau}$ is non-zero. 
If the expectation value $z_{\alpha\tau}$ drops to zero, the corresponding orbital $\alpha\tau$ loses the quasiparticle weight and becomes Mott insulating, and its orbital-resolved filling factor is also enforced to be half-filled. In the numerical calculation in the main text, we also assumed the expectation value $z_{\alpha\tau}$ is irrelevant to the valley (spin) index.

\begin{figure}[t]
    \centering
    \includegraphics[width=\linewidth]{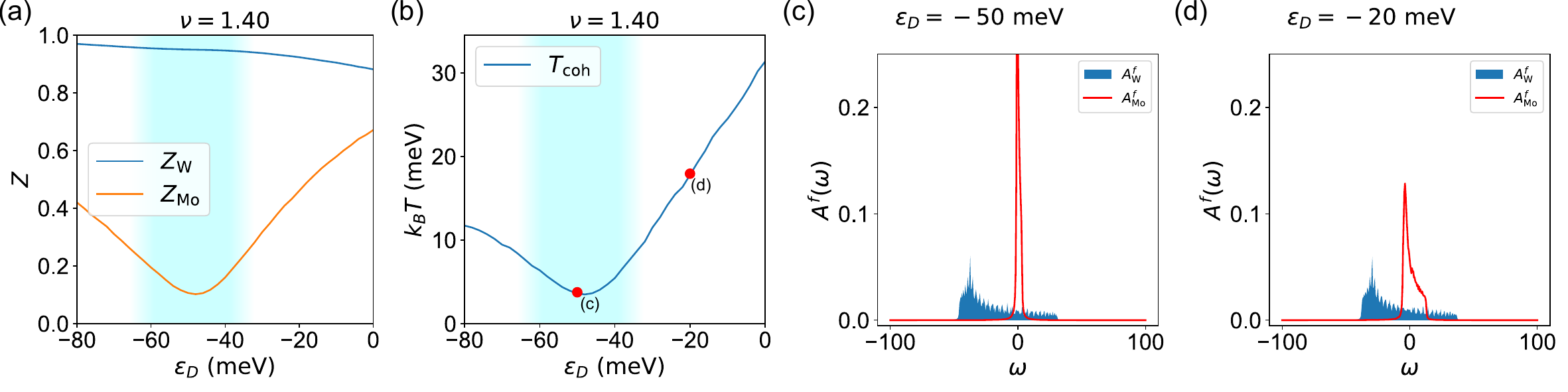}
    \caption{(a) The quasiparticle weights of both the $c$ and $d$ orbitals as functions of the displacement field with fixed total hole filling factor $\nu = 1.4$. (b) The coherent temperature scale $T_{\rm coh}$ estimated by slave fermion local density of states with fixed total hole filling factor $\nu = 1.4$. Note that this figure is the same as Fig.~\ref{fig:phase_diagrams}(e) in the main text. (c) The slave fermion local density of states for both orbitals with displacement field $\varepsilon_D = -50\rm \,meV$, at which the $d$ orbital filling factor is very close to $\nu_{\rm Mo} \approx 1$. (d) The slave fermion local density of states with displacement field $\varepsilon_D = -20\,\rm meV$, at which the $d$ orbital filling factor $\nu_{\rm Mo}$ is away from half-filling. In these calculations we choose the on-site interaction strength $U = 70\rm\,meV$.}
    \label{fig:add-Tcoh}
\end{figure}

\section{Coherence temperature}\label{appsec:Tcoh}

We consider the coherence temperature scale $T_{\rm coh}$ in the heavy Fermi liquid state.
This temperature characterizes the finite temperature onset of the initial Kondo screening. It can be estimated in terms of the effective bandwidth of the $d$ orbitals in the saddle-point calculation at zero temperature.
Hence, a practical way \cite{kourris_kondo_2023} of estimating $T_{\rm coh}$ is using the inverse of the local density of states of the slave fermion: $k_BT_{\rm coh} = 1/A^f_{\rm Mo}(\omega = 0)$, in which $A^f_{\rm Mo}(\omega)$ corresponds to the ``spectral function'' of the $d$ orbital slave fermion. 
Note that it is different from the local density of states of the physical fermions $d^\dagger$, which will be discussed in App.~\ref{appsec:ldos}. 
Since the slave fermion Hamiltonian $H^f$ is a quadratic fermionic Hamiltonian, the local density of states of slave fermions can be easily obtained via the inverse of its ``Bloch Hamiltonian'':
\begin{align}
    A^f_{\alpha\tau}(\omega) &= -\frac{1}{\pi}{\rm Im}\,\frac{1}{N}\sum_{\vk}\left(\frac{1}{\omega + i0^+ - h^f(\vk)}\right)_{\alpha\tau,\alpha\tau}\,,\\
    h^f(\vk) &= \left(
    \begin{array}{cc}
        h^f_{+}(\vk) &  \\
        & h^f_-(\vk)
    \end{array}
    \right)\,,\\
    h^f_\tau(\vk) &= \left(
    \begin{array}{cc}
        Z_{\rm Mo} \epsilon^\tau_d(\vk) - \lambda_d + \lambda^0_d -\frac{\varepsilon_D}{2} - \mu & \sqrt{Z_{\rm Mo}Z_{\rm W}}\epsilon_{cd}(\vk) \\
        \sqrt{Z_{\rm Mo}Z_{\rm W}}\epsilon^*_{cd}(\vk) & Z_{\rm W} \epsilon^\tau_c(\vk) - \lambda_c + \lambda^0_c + \frac{\varepsilon_D}{2} - \mu
    \end{array}
    \right)\,,
\end{align}
in which $\epsilon^\tau_d(\vk), \epsilon^\tau_c(\vk)$, and $\epsilon_{cd}(\vk)$ are the Fourier transformation of the real space hoppings $t_{d}e^{i\tau\phi^d_{ij}}, t_{c}e^{i\tau\phi^c_{ij}}$ and $t_{cd}$, respectively. 

In particular, if we assume the $d$ orbitals are exactly flat with vanishing bandwidth [$\epsilon_d(\vk) \rightarrow 0$] and the $c$ orbitals are non-interacting ($Z_{\rm W} = 1$), the slave fermion Hamiltonian can be approximately written in the following form:
\begin{equation}
    h^f_\tau(\vk) \sim \left(
        \begin{array}{cc}
            0 & \sqrt{Z_{\rm Mo}} \epsilon_{cd}(\vk) \\
            \sqrt{Z_{\rm Mo}} \epsilon^*_{cd}(\vk) & \epsilon^\tau_c(\vk) + \mathcal{E}_c
        \end{array}
    \right)\,.
\end{equation}
Here $\mathcal{E}_c$ contains the contributions from Lagrange multipliers $\lambda_c, \lambda^0_c$, chemical potential $\mu$ and displacement field $\varepsilon_D$, such that the dispersion of $\epsilon_c^{\tau}(\vk) + \mathcal{E}_c$ crosses the Fermi level with the correct filling factor. We assume the off-diagonal elements are perturbative, and the two eigenvalues of the matrix $h^f_\tau(\vk)$ are:
\begin{align}
    \omega_+(\vk) &\sim \epsilon_c(\vk) + \mathcal{E}_c \,,\\
    \omega_-(\vk) &\sim - \frac{Z_{\rm Mo}|\epsilon_{cd}(\vk)|^2}{\epsilon_c(\vk) + \mathcal{E}_c}\,,
\end{align}
in which the dominant component of the eigenvector for $\omega_-(\vk)$ corresponds to the $d$ orbital. Thus, charge excitations that exhibit a significant overlap with the $d$ orbital are predominantly distributed within an energy interval with a width denoted as $k_B T_{\rm coh} \sim \frac{Z_{\rm Mo} t_{cd}^2}{D_{\rm W}}$, where $D_{\rm W}$ is the bandwidth of the conduction band. This expression resembles the definition of the coherence temperature scale $T^* \sim {r^2}/{D_c}$ introduced in Ref.~\cite{burdin_coherence_2000}. 

However, the realistic Hamiltonian of the hetero-bilayer TMDC does not have a vanishing $d$ orbital dispersion. In contrast, the bandwidth of the $d$ orbital $D_{\rm Mo}$ is much larger than the hybridization $t_{cd}$. If the bandwidth of the heavy band is taken into account, the coherent temperature scale has to be estimated from the following matrix:
\begin{equation}
    h^f_\tau(\vk) \sim \left(
        \begin{array}{cc}
            Z_{\rm Mo}\epsilon^\tau_{d}(\vk) + \mathcal{E}_d & \sqrt{Z_{\rm W}Z_{\rm Mo}} \epsilon_{cd}(\vk) \\
            \sqrt{Z_{\rm W}Z_{\rm Mo}} \epsilon^*_{cd}(\vk) & Z_{\rm W}\epsilon^\tau_c(\vk) + \mathcal{E}_c
        \end{array}
    \right)\,.
\end{equation}
If the hybridization terms in the off-diagonal elements are still treated as perturbations, the eigenvalue that is dominated by the $d$ orbital will be:
$$
    \omega_-(\vk) \sim Z_{\rm Mo}[\epsilon_d(\vk) + \mathcal{E}_d] - \frac{Z_{\rm Mo}Z_{\rm W}|\epsilon_{cd}(\vk)|^2}{\epsilon_c(\vk) + \mathcal{E}_c}\,.
$$
Consequently, the coherent temperature scale is different from the case with exact flat heavy bands, and it can be estimated as follows:
\begin{equation}
    k_BT_{\rm coh} \sim Z_{\rm Mo} D_{\rm Mo} + \frac{Z_{\rm Mo}Z_{\rm W}t_{cd}^2}{D_{\rm W}}\,.
\end{equation}
When compared with the exact flat band case $t_d = 0$, it is significantly increased by the bandwidth of the $d$ orbital.

By calculating the local density of states of the slave fermions, we are able to get the coherence temperature with different displacement field strength values. 
The results can be found in Fig.~\ref{fig:add-Tcoh}(b). 
Observing the plot, it becomes evident that $T_{\rm coh}$ closely resembles the curve of the quasiparticle weight multiplied by a factor that is approximately equal to the $d$ orbital bandwidth, and $T_{\rm coh}$ in the heavy Fermi liquid region is strongly suppressed to $k_B T_{\rm coh} \sim 5\,\rm meV$. 
In contrast, when the $d$ orbital filling factor is doped far away from $\nu_{\rm Mo} = 1$, the heavy fermion quasiparticle weight gets large and thus the coherent temperature can approach $T_{\rm coh}\gtrsim 100\,\rm K$. 

As a reference, we also show the slave fermion local density of states for both orbitals at two points in the phase diagram in Figs.~\ref{fig:add-Tcoh}(c-d). 
The local density of states in Fig.~\ref{fig:add-Tcoh}(c) is obtained with $\varepsilon_D = -50\,\rm meV$ and $\nu = 1.4$, which is in the heavy Fermi liquid region, while in Fig.~\ref{fig:add-Tcoh}(d), the local density of states is obtained outside of the heavy Fermi liquid region. 
The heavy Fermi liquid state has a narrower heavy band and a higher slave fermion density of states at $\omega = 0$, and thus a much lower $T_{\rm coh}$. 

\section{Local density of states}\label{appsec:ldos}

The local density of states shown in the previous appendix is obtained from the slave fermion operators $f^\dagger_{\hat{c}}$ or $f^\dagger_{\hat{d}}$, instead of the physical fermions $c^\dagger$ or $d^\dagger$. Local charge fluctuation is not considered, and hence the lower and upper Hubbard bands are not captured. To correctly describe the local charge fluctuation, slave spin excited states need to be considered. 

We write the physical fermion operators as the product of slave fermion operator and slave spin operator, and we write the many-body eigenstates as tensor products of the local slave spin eigenstates and Slater determinants of slave fermion states.
Hence, the local density of states of the physical degrees of freedom can be obtained from the following expression, which has been derived in Ref.~\cite{yu_mott_2011}:
\begin{align}
    A_{\alpha\tau}(\omega) = & \frac{1}{N}\sum_{\vk}\sum_m\Bigg{(}\sum_{i, \epsilon_{\vk,i} > 0}\delta(\omega + E_g - E_m - \epsilon_{\vk, i}) |u_{\alpha\tau, i}(\vk)|^2 \frac{|\langle m | S^+_{\alpha\tau} | g \rangle|^2}{n_{\alpha\tau}(1 - n_{\alpha\tau})} \mathcal{F}^+_{\alpha\tau, m}  \nonumber \\
    & + \sum_{i, \epsilon_{\vk, i} < 0}\delta(\omega - E_g + E_m - \epsilon_{\vk,i}) |u_{\alpha\tau, i}(\vk)|^2 \frac{|\langle m | S^-_{\alpha\tau} | g \rangle|^2}{n_{\alpha}(1 - n_{\alpha\tau})} \mathcal{F}^-_{\alpha\tau, m} \Bigg{)}\,, \label{eqn:def-ldos} \\
    \mathcal{F}^+_{\alpha\tau, m} =& \left\{
    \begin{array}{lr}
        1 & m = g  \\
        \frac{1 - Z_{\alpha\tau}}{n_{\alpha\tau}^{-1} - Z_{\alpha\tau}} &  m \neq g
    \end{array}
    \right.\,,\\
    \mathcal{F}^-_{\alpha\tau, m} =& \left\{
    \begin{array}{lr}
        1 & m = g  \\
        \frac{1 - Z_{\alpha\tau}}{(1 - n_{\alpha\tau})^{-1} - Z_{\alpha\tau}} &  m \neq g
    \end{array}
    \right.\,.
\end{align}
Here we use $E_m$ and $|m\rangle$ to represent the eigenvalues and eigenstates of the local slave spin Hamiltonian $H_i^S$, and the summation over $|m\rangle$ includes all the eigenstates of $H_i^S$. Specifically, the ground state of $H_i^S$ is denoted by $|g\rangle$. We also use $\epsilon_{\vk,i}$, $u_{\alpha\tau,i}(\vk)$ to represent the $i$-th eigenvalue and eigenvector of the slave fermion Hamiltonian $h^f(\vk)$. The factors $\mathcal{F}^\pm_{\alpha\tau}$ guarantee that Eq.~(\ref{eqn:def-ldos}) satisfy the following sum rules of the spectral functions:
\begin{align}
    \int_{-\infty}^{\infty}d\omega A_{\alpha\tau}(\omega) &= 1\,,\\
    \int_{-\infty}^0 d\omega A_{\alpha\tau}(\omega) &= n_{\alpha\tau}\,.
\end{align}

\begin{figure}[t]
    \centering
    \includegraphics[width=0.67\linewidth]{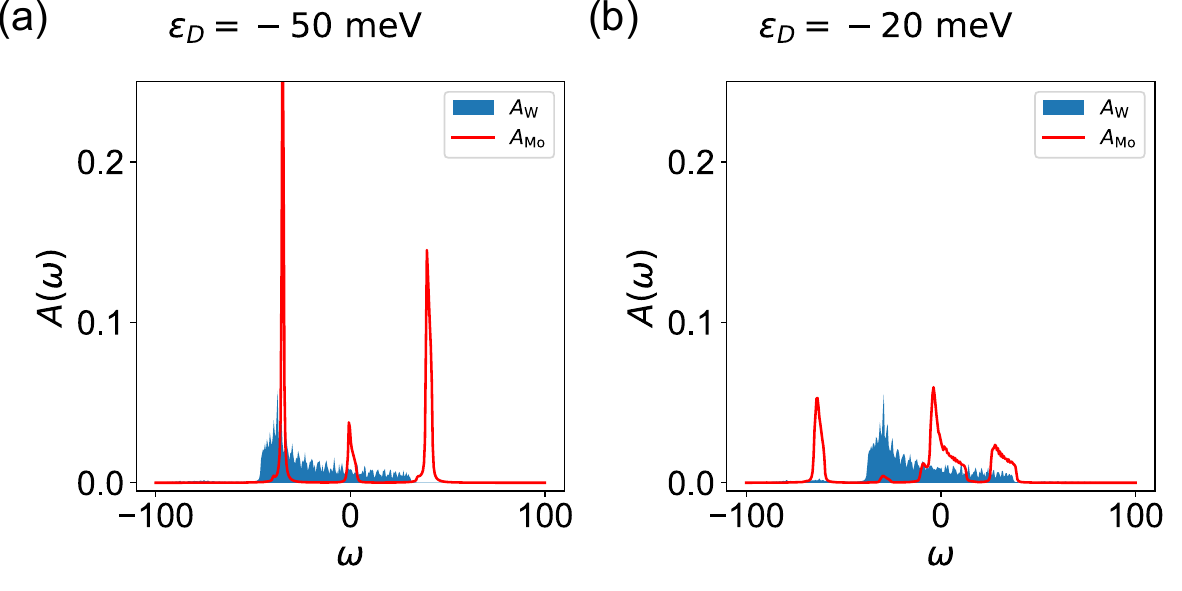}
    \caption{The local density of states of the physical fermions with different displacement field. The parameters are chosen to be the same as in Figs.~\ref{fig:add-Tcoh}(c-d).}
    \label{fig:add-ldos}
\end{figure}

Using the self-consistent solutions of the slave spin and slave fermion Hamiltonians, we numerically evaluated the local density of states for both the $c$ and $d$ orbitals with the same parameters as in Figs.~\ref{fig:add-Tcoh}(c-d). The results can be found in Fig.~\ref{fig:add-ldos}. 
In both the heavy Fermi liquid state (a) and normal Fermi liquid state (b), the spectral peaks of the $d$ orbital near $\omega = 0$ are lower than the peak in the slave fermion local density of states, which is a consequence of a small $Z_{\rm Mo}$. 
The incoherent upper and lower Hubbard bands separated by $U$ are clearly visible in both cases. 
Since the bare bandwidth of the $d$ orbital is not negligible when compared with $U$, the coherent peak of $A_{\rm Mo}(\omega)$ near the Fermi energy starts getting wider noticeably when the upper Hubbard band moves close to $\omega = 0$, even if it is still obviously above $\omega = 0$, as seen in Fig.~\ref{fig:add-ldos}(b). 
This indicates a large quasiparticle weight and a reduced effective mass for the heavy fermion.
As a result, the width of the heavy Fermi liquid region along the displacement field potential axis $\Delta \varepsilon_D$, in which the quasiparticle weight of the $\rm Mo$ orbital remains very small ($Z_{\rm Mo} \ll 1$), will be narrower than the on-site interaction $U$.

\section{Orbital-selective Mott transition via strong interaction}\label{appsec:osmp-U}

\begin{figure}[t]
    \centering
    \includegraphics[width=\linewidth]{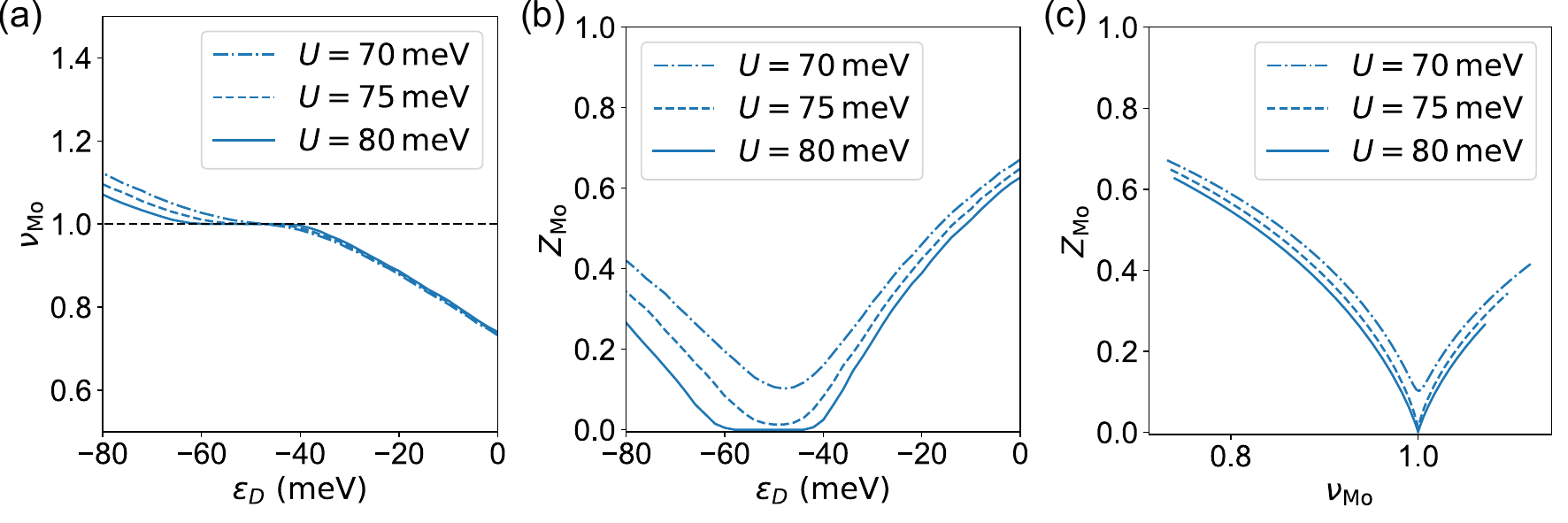}
    \caption{(a) The filling factor of the $d$ orbital as functions of displacement field strength with different interaction strength $U = 70 \rm\,meV$, $U = 75 \,\rm meV$ and $U = 80\rm\, meV$. (b) The quasiparticle weight of the $d$ orbital with different interaction strength. (c) The $d$ orbital quasiparticle weight as functions of its filling factor with different interaction strength. The total filling factor is set to $\nu = 1.4$.}
    \label{fig:add-U-osmp}
\end{figure}

In the main text, we discussed the orbital-selective Mott phase obtained by reducing the conduction electron density. The orbital-selective Mott phase transition can be achieved by increasing the interaction strength as well. The $d$ orbital filling factor at the same total filling factor $\nu = 1.4$ with different interaction strength up to $U = 80\rm\,meV$ is shown in Fig.~\ref{fig:add-U-osmp}(a). It is obvious that the ``plateau,'' in which the heavy Fermi liquid state is situated, broadens as the interaction becomes stronger. However, the width of the ``plateau'' along the $\varepsilon_D$ axis is still much smaller than the on-site interaction $U$. The quasiparticle weight of the $d$ orbital, shown in Figs.~\ref{fig:add-U-osmp}(b) and \ref{fig:add-U-osmp}(c), also exhibits the orbital-selective Mott phase at total filling $\nu = 1.4$ when the interaction is increase to $U = 80\rm\,meV$. 

\section{Phase diagram in experiment}\label{appsec:exp}

\begin{figure}
    \centering
    \includegraphics[width=0.5\linewidth]{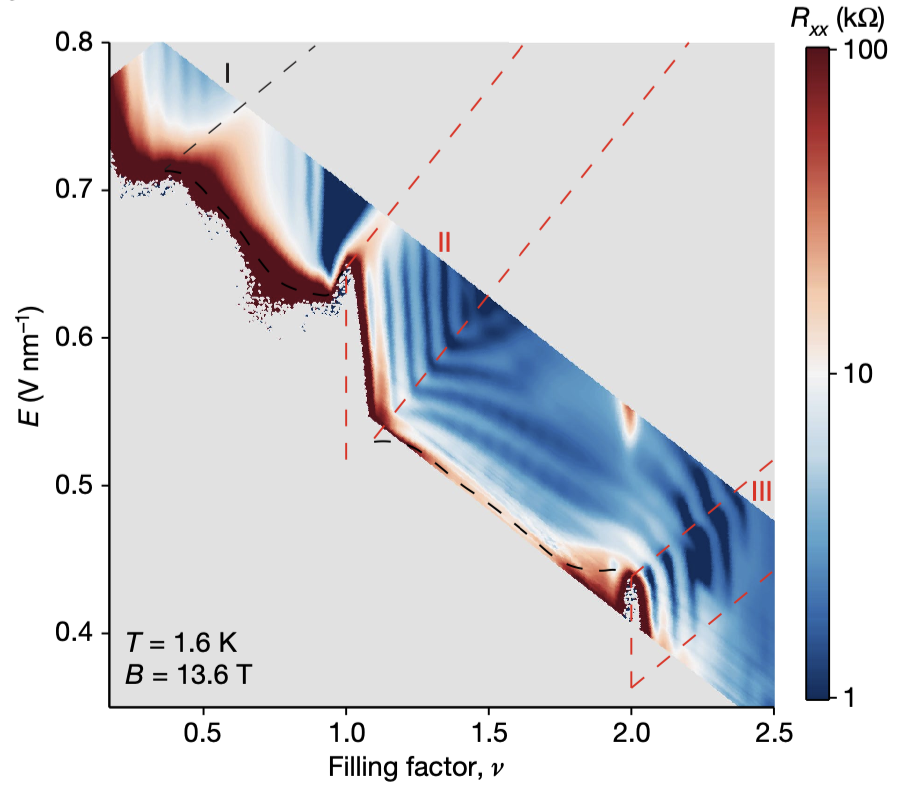}
    \caption{The experimental electrostatics phase diagram in the parameter space $(\nu, E)$, in which $\nu$ is the total electron density, and $E$ is the electric field strength. The heavy Fermi liquid regime with filling factor $\nu = 1 + x$ is labeled by ``II''. This figure is reproduced from Fig.~2(a) in Ref.~\cite{zhao2022gate}.}
    \label{fig:exp-phase-diagram}
\end{figure}

In Fig.~\ref{fig:exp-phase-diagram} we show the experimentally observed electrostatics phase diagram adapted from Ref.~\cite{zhao2022gate}. The two axes of this phase diagram are the total filling factor $\nu$ and the displacement field strength $E$, which effectively is a linear function to its corresponding potential $\varepsilon_D$. The measurement is performed with the presence of an out-of-plane magnetic field. Therefore, quantum oscillation patterns can be observed when $\nu$ is varied. The Landau fans are vertical within the regions labeled by red dashed lines, which indicate the Fermi surfaces are almost unchanged by the displacement field when $\nu$ is fixed. 
This phenomenon indeed corresponds to the plateaus in the filling factors of the two orbitals in Fig.~\ref{fig:phase_diagrams}(b), marking the formation of the heavy Fermi liquid state. 
We also note that the shape of this region in the $(\nu, E)$ phase diagram is qualitatively identical to Fig.~\ref{fig:phase_diagrams}(a). 
This phase diagram also shows that, by controlling the total filling factor $\nu$ and the displacement field $E$ together, both $\nu_{\rm Mo}$ and $\nu_{\rm W}$ can be indirectly tuned in experiment.

\end{document}